\documentclass[12pt]{article}
\usepackage{amsfonts,amsmath}
 \usepackage{amssymb}
\usepackage[hypertex] {hyperref}

\makeatletter \@addtoreset{equation}{section} \makeatother

\newcommand{\husp}{{h_{usp}}}
\newcommand{\ho}{{h_{o}}}
\newcommand{\musp}{{m_{usp}}}
\newcommand{\mo}{{m_{o}}}
\newcommand{\hu}{{h_{u}}}

\newcommand{\cV}{{\mathcal V}}

\newcommand{\JJJ}{ \mathcal{J}}

\newcommand{\V}{{\cal V}}

\newcommand{\be}{ \begin{equation}}

\newcommand{\II}{{\mathcal I}}
\newcommand{\U}{{\mathcal U}}

\newcommand{\ee}{\end{equation}}
\newcommand{\bee}{\begin{eqnarray}}
\newcommand{\beee}{\begin{array}}
\newcommand{\eee}{\end{eqnarray}}
\newcommand{\eeee}{\end{array}}

\newcommand{\cc}{{\cal C}}

\newcommand{\ga}{\alpha}

\newcommand{\gb}{\beta}

\newcommand{\gga}{\gamma}

\newcommand{\rhs}{{\it r.h.s.} }

\newcommand{\ie}{{\it i.e.,} }
\newcommand{\ls}{\!\!\!\!\!\!}

\newcommand{\gs}{\sigma}

\newcommand{\go}{\omega}

\newcommand{\q}{\,,\qquad}

\newcommand{\nn}{\nonumber}
\newcommand{\mmu}{m_u}

\newcommand{\half}{\frac{1}{2}}

\newcommand{\p}{\partial}

\newcommand{\f}{\frac}
\newcommand{\C}{{\cal C}}

\newcommand{\hs}{{hs}}

\newsavebox{\ver}
\newsavebox{\verp}
\newsavebox{\gorp}
\newsavebox{\toch}

\begin{document}

\begin{flushright}
{\small FIAN/TD/26-12}
\end{flushright}
\vspace{1.7 cm}

\begin{center}
{\large\bf  Multiparticle extension of the higher-spin algebra}

\vspace{1 cm}

M.A.~Vasiliev
 \vglue 0.3  true cm

 I.E.Tamm Department of Theoretical Physics, Lebedev Physical
 Institute,\\
 Leninsky prospect 53, 119991, Moscow, Russia

\end{center}

\vspace{0.4 cm}

\begin{abstract}
\noindent
Multiparticle extension of a higher-spin  algebra $l$ is introduced as the
Lie superalgebra associated with the universal enveloping algebra $U(l)$. While
conventional higher-spin symmetry does not mix $n$-particle states with
different $n$,  multiparticle symmetries do so. Quotients of multiparticle
algebras  are considered, that act on the space of $n$-particle states with
$0\leq n\leq k$ analogous to the space of first $k$ Regge trajectories of
String Theory. Original  higher-spin algebra is reproduced at $k=1$. Full
multiparticle algebras are conjectured to describe vacuum symmetries of
string-like extensions of higher-spin gauge theories. Relation of the
multiparticle algebras with $3d$ current operator algebras is described.
The central charge parameter, to be related to the parameter ${\mathcal N}$
in  $AdS/CFT$ correspondence, enters via the definition of supertrace.
Extension to higher $p$-brane-like symmetries is introduced inductively.

\end{abstract}

\newpage
\tableofcontents

\newpage

\sbox{\ver}{\line(0,1){4}} \sbox{\gorp}{\line(1,0){7}}
\sbox{\verp}{\line(0,1){7}} \sbox{\toch}{\circle*{1}}

\section{Introduction}

Higher-spin (HS) gauge theories  describe interactions of massless
fields of all spins. The first example of full nonlinear HS theory
was given in the $4d$ case \cite{con}, while its modern formulation
was worked out  in \cite{more} (see \cite{Vasiliev:1999ba} for a
review). HS  gauge theories involve infinite towers of massless
(gauge) fields of higher spins. In this respect HS gauge theory is
analogous to String Theory which also describes interactions of
excitations of states of all spins. These two classes of theories
are however different in several respects.

Known HS gauge theories only involve totally symmetric
fields while String Theory  contains HS fields of various symmetry types.
Field spectra of HS gauge theories are somewhat analogous to the first
Regge trajectory of String Theory, though describing only massless (gauge)
 fields of higher spins. On the other hand, String Theory describes
only massive  higher spins. Another distinction is that
HS theories admit consistent interactions only in a curved background
which is $(A)dS_d$ in the most symmetric case, while fully consistent
formulation of String Theory is available in a Ricci flat background.

It was anticipated for a long time that HS theory and String Theory
should  be  related and, eventually, String Theory should be
understood as some HS theory where masses are generated via spontaneous
breakdown of HS symmetries (for example, this was conjectured  in
\cite{Vasiliev:1987zv}). Although this conjecture  is supported by the analysis of
high-energy limit of string amplitudes \cite{Gross:1988ue} and passed some
nontrivial checks \cite{Bianchi:2003wx,Beisert:2004di,Bianchi:2004ww,Bianchi:2005yh},
no satisfactory understanding of this relation beyond the free field sector
of the tensionless limit of String Theory
\cite{Lindstrom:2003mg,Bonelli:2003kh,Sagnotti:2003qa} was available.
An interesting  idea of singleton string whose spectrum is represented
by multiple tensor products of singletons was put forward  in
\cite{Engquist:2005yt,Engquist:2007pr}. Somewhat similarly,
it was recently conjectured \cite{Chang:2012kt} that String Theory should
admit an interpretation of a theory of bound states of HS gauge theory.
Consideration of this paper agrees with these conjectures, specifying a
symmetry underlying a string-like extension of HS gauge theory.

More in detail, it is anticipated  that there should exist a
string-like HS gauge theory that describes at least as many degrees
of freedom as String Theory. However, as long as HS symmetries are
unbroken, this hypothetical theory will be free of an $\ga'$-like
mass scale parameter, describing massless fields along with massive
fields whose masses are scaled in units of inverse $AdS$ radius.
Similarly to usual HS gauge theory, if exists, such string-like HS
gauge theory could only be formulated in a curved background which
is $(A)dS$ in the most symmetric case. Upon HS symmetries are
spontaneously broken, the theory should acquire an independent mass
scale parameter giving mass to HS fields, and admit formulation in
flat space. Involving infinite tower of massive HS fields, in this
case it should reduce to one or another version of String Theory. As
in usual HS theory, the key step consists of identification of a
global symmetry gauging of which underlies a string-like HS gauge
theory. As discussed below, such a symmetry has to obey a number of
nontrivial conditions. The multiparticle algebra analyzed in this
paper will be argued to pass these conditions hence providing a
promising candidate for the HS algebra of a string-like HS theory.
Being broken in a vacuum associated with usual String Theory,
multiparticle symmetry can hardly be seen in this framework since
most of related transformations will have Stueckelberg form. In
other words, to uncover hidden symmetries of String Theory one has
to find a string-like HS gauge model where these symmetries are
unbroken (and, hence,
 related states are massless). Identification of  multiparticle  algebra as a candidate
 for such a symmetry  is one of main outputs of this paper.

Naively, the difference between the two classes of theories is
minor. HS theories are formulated in terms of  fields
$B(Y|X)$ that depend on space-time coordinates $X$ and auxiliary
variables $Y^A$. The latter, depending on a model, can be either spinors
\cite{Ann,more} or a pair of vectors \cite{Vasiliev:2003ev}. The
variables $Y^A$ are noncommutative, obeying commutation relations
\be
\label{weyl1}
[Y^A\,,Y^B] = 2C^{AB}\,
\ee
with some non-degenerate antisymmetric matrix
$C^{AB}$ which is either the charge conjugation matrix with spinor
indices $A, B$ or  has the form $C^{AB}=\epsilon^{\ga\gb}\eta^{ab}$
with $A=(a,\ga)$ where $\ga=1,2$, $\epsilon^{\ga\gb}=-\epsilon^{\gb\ga}\neq 0$,
$a$ is the vector index of
$o(d-1,2)$ and $\eta^{ab}$ is an $o(d-1,2)$ invariant metric. These
oscillators are analogous to a pair of string oscillators, say
$x^n_1$ and $x^m_{-1}$, that satisfy
\be [x_1^n\,,x^m_{-1}] =
\eta^{nm}\,,
\ee
where $\eta^{nm}$ is Minkowski metric in $d$
dimensions.

It looks like it is enough to let more species of oscillators
$Y^A\to Y^A_i$ ($i=1,\ldots r$) be present to get HS theory
closer to String Theory picture which should emerge in the
 $r\to \infty$ limit. This idea is
supported by the analysis of unfolded formulation of free mixed
symmetry HS gauge fields  which were shown
\cite{Alkalaev:2003qv,Skvortsov:2008vs,Boulanger:2008up,Skvortsov:2009zu}
to be naturally described in terms of differential forms
$\go(Y_i|x)$ of various degrees, valued in appropriate tensor
$o(d-1,2)$-modules  realized by polynomial functions $\go(Y_i|x)$ of
oscillators $Y_i^A$. However, to go beyond the free field level, it
is  necessary to find such a non-Abelian algebra $\hs$ that fields
of the unfolded formulation of the theory fit into  $hs$-modules. In
particular, 1-forms among $\go(Y_i|x)$ should be valued in its
adjoint representation. A strong  criterium, called admissibility
condition  \cite{KV0},  requires a HS algebra to admit a unitary
module that decomposes into direct sum of unitary modules of the
space-time symmetry algebra $s$, whose pattern matches the list of
relativistic fields associated with the list of forms $\go(Y_i|x)$.
For symmetric HS fields this is indeed the case
\cite{KV0,Konstein:ij} due to  Flato-Fronsdal theorem \cite{FF} and
its higher-dimensional generalization \cite{Vasiliev:2004cm} which
relates tensor product of a pair of scalar and/or spinor unitary
modules of the conformal algebra $o(d-1,2)$ in $d-1$ dimensions to
the towers of  massless fields in $d$ dimensions. In this
realization, the $AdS_d$ HS algebra  is identified with the algebra
of endomorphisms of the space of single-particle states of conformal
fields in $d-1$ dimension. However, no analogue of this construction
appropriate for the description of mixed symmetry fields of general
type is available. The problem is most obvious for odd $d$ where
conformal scalar and spinor fields are the only  unitary propagating
conformal fields. This indicates that generalization of HS theory to
mixed symmetry fields and, eventually, to String Theory, may require
some deviation from the standard constructions of HS theory.

1-forms $\go(Y|x)$ valued in $s\subset \hs$, describe
vielbein and connection of the spin two gravitational field. In the
usual HS theory of symmetric fields, the corresponding gravitational
fields were associated with the (sub)algebra of bilinears of the oscillators
\be
T^{AB}=\half \{Y^A\,, Y^B\}\,.
\ee
For instance, in the $4d$ spinor realization where $A,B=1,\ldots 4$, $T^{AB}$
are generators of the $AdS_4$ algebra $sp(4)\sim o(3,2)$. For arbitrary dimension
$d$, generators of $o(d-1,2)$ were identified \cite{Vasiliev:2003ev} with
the subalgebra of $sp(2(d+1))$ spanned by those $T^{AB}$ that are invariant under
the $sp(2)$ subalgebra rotating indices $\ga$ of $Y^{a \ga}$.

However  straightforward extension of this construction to any number
$r$ of oscillators
\be
\label{N}
T^{AB}=\half \sum_{i=1}^r \{Y_i^A\,, Y_i^B\}\,
\ee
does not respect the admissibility condition facing the following problem.
A natural framework for  unitary $hs$-modules is
provided by tensor products of Fock modules where the oscillators $Y^A_i$ act.
Let $E$ be the energy operator among $T^{AB}$. If the lowest energies for spin
$s$ fields were $E(s)$ in the $r=1$ $\hs$-module (recall that
in the absence of a free mass parameter, lowest energies in $AdS_d$ are scaled
in units of the inverse $AdS$ radius), in the tensor product of $r$
such modules energies will increase like  $r E(s)$. Since the lowest energy
determines the mass of a particle, if it described a massless symmetric
field in the $r=1$ case, it will correspond to certain massive (and hence
non-gauge) field at higher $r$. In particular, spin
one and  two fields become massive at $r>1$, \ie
the resulting theory can  contain neither Yang-Mills theory, nor gravity.

So far,  no non-Abelian HS algebra appropriate for description of general
mixed symmetry fields was available, though some
particular mixed symmetry fields result from gauging of
the HS algebra associated with the tensor product of fermions in any dimension
\cite{Vasiliev:2004cm} as well as with $4d$ conformal HS algebras of \cite{FLA,FL} and
their further $4d$ \cite{Vasiliev:2001zy,Sezgin:2001yf,Alkalaev:2002rq} (for the
respective Flato-Fronsdal like theorem see \cite{Boulanger:2011se})
and higher-dimensional \cite{Sezgin:2001ij,Bekaert:2009fg}  extensions.

A well-known feature of  conventional  formulation of String Theory,
which seems to be closely related to the above discussion of HS theory, is
 that its consistent generalization to $AdS$ background is far from being
 trivial. Indeed, as in any  relativistic theory, Lorentz symmetry acts
 on all space-time (spinor-)tensors  both in HS theory and
in String Theory. Hence, Lorentz generators should have a form
(\ref{N}) where summation is over all modes that carry space-time indices.
That commutator of translations (transvections) in $AdS$ algebra gives
Lorentz generators requires the $AdS$ translation generators to be built from
all modes. However, as in  HS theory, this would immediately lead to
wrong (infinite) vacuum energy of graviton. Hence,
 translation generators in String Theory are built
solely from zero modes, which construction
admits no $AdS$ deformation (see however interesting work \cite{Polyakov:2011sm}
where an extended formulation of String Theory, that avoids this
problem, was proposed). This feature of String Theory
indicates that the straightforward  construction via tensoring of
oscillators is too naive in the both cases.

In this paper, we propose a class of  algebras that extend
usual HS algebras in a String Theory fashion, avoiding the most obvious problems
mentioned above.
The proposed construction was deduced from the analysis of \cite{Gelfond:2013xt}
of current operator algebra of $3d$ massless free theory, which generates
symmetries  of the space of multiparticle states of the
$AdS_4$ HS theory and its boundary image. Hence, we call them multiparticle
algebras. The purely algebraic approach of this paper provides an
efficient tool for the description  of the current operator
algebra of \cite{Gelfond:2013xt}, leading to manifest formulae
for the current OPE in Section \ref{Current operator algebra}. One of the
surprising outputs of this  construction is that OPE's of currents with different
number ${\mathcal N}$ of constituent free
fields are described by different basis choices in the same multiparticle
algebra. This  may look  surprising since this property has no analogue in two-dimensional
conformal field theory where ${\mathcal N}$ affects the central charge of
Virasoro algebra. Indeed, that central extension in Virasoro algebra is
nontrivial literally means that models with different central charges do not
correspond to different bases  of Virasoro algebra. As follows from our
consideration, the situation in higher-dimensional conformal theories is different.
In particular, multiparticle algebras admit no nontrivial central
extension. Practically, the respective bases are uniquely fixed by the conditions
that (i) Wick theorem, is respected as the characteristic property of free field
theory and (ii) the central term  properly depends on  ${\mathcal N}$. Since
different free theories are described by the same multiparticle algebra, it is
appealing to speculate that,  using the same multiparticle algebra, it may
even be possible to formulate nonlinear conformal systems not respecting Wick theorem
(see also Conclusion).

 It should be stressed that our construction applies to a very general class
of theories including the  $4d$ $N=4$ SYM boundary theory closely related to
conventional Superstring Theory. Similarly, the analysis of current operator
algebra of  \cite{Gelfond:2013xt} goes beyond the $3d$ case, allowing in particular
to evaluate $n$-point functions of $4d$ conformal currents.

 The formal definition is simple.
Let a HS algebra  $\hu (V)$ be the Lie algebra of maximal symmetries of
$V$, \ie of the free theory of fields $\Phi$ that have $V$ as the space
of single-particle states. Multiparticle algebras $\mmu(V)$ are
appropriate real forms of the complex Lie superalgebras
 associated with the universal enveloping algebras $ U(\hu(V))$.
 Multiparticle algebra
acts on the space of all multiparticle states
in a theory where $V$ is the space of single-particle states.

Algebras $\mmu^k(V)$ act on the space of $r$-particle states with $r\leq k$.
 $\hu(V)\subset \mmu^k (V)$ for any $k\geq 1$.
In particular, $\mmu^1(V)=\hu(V)\oplus u(1)$ where $u(1)$
represents the symmetry of the physical vacuum which is the space of 0-particle states.
Algebras $\mmu^k(V)$  are certain quotients
of $\mmu(V)$  and should be associated with a
theory which roughly speaking describes first $k$ Regge trajectories
of String Theory. $\mmu(V)$ should be associated with the full-fledged
string-like extension of HS theories.
We believe that the proposed scheme has a potential to unify String Theory and HS
theory within a theory which contains both of them as different particular cases
and/or limits.

The paper is organized as follows.
In Section \ref{HSA} general structure of known HS algebras is recalled.
In Section \ref{genw} we present  construction of associative multiparticle
algebra which is illustrated in Section \ref{weyl} by the example
of  Weyl algebra. It is applied to description of current operator
algebras of \cite{Gelfond:2013xt} in Section \ref{Current operator algebra} and
to extension of every HS algebra $\hu(V)$ to multiparticle algebras $\mmu(V)$ and
$\mmu^k(V)$ in Section \ref{hssa}, where further generalizations of multiparticle algebras
to be associated with $p$-brane extensions of String Theory are also introduced.
In Conclusion, some properties of yet hypothetical string-like HS theory
are briefly discussed as well as possible extension of the obtained results
to non-free current operator algebras.

\section{Higher-spin  algebras}
\label{HSA}

From the perspective of bulk HS gauge theories in $AdS$,
HS algebras represent global symmetry   of a maximally symmetric vacuum
solution of the nonlinear HS gauge theory in question. They should be
distinguished from local HS symmetries of HS gauge theories,  resulting from gauging
(localization) of global $HS$ symmetries along with their further field-dependent
deformation.\footnote{The latter phenomenon is typical for any theory of gravity where
diffeomorphisms can be interpreted as a deformation of localized Poincar{\' e}
or $(A)dS$ transformations by curvature-dependent terms
\cite{Utiyama,kibble,chwest,MM,PVN}.} In this paper, we
focus on the  global multiparticle  algebras, which provide
starting point for the search of the full-fledged nonlinear multiparticle gauge theories.

All HS algebras underlying known nonlinear HS gauge theories admit the following realization.
Let $V_\Phi$ be the space of single-particle states of a set of free unitary conformal fields
$\Phi$. We will use notation $H(V_\Phi)$ for the complex associative algebra of
endomorphisms $End (V_\Phi;\mathbb{C})$. As such,
$H(V_\Phi)$ is closely related to the algebra of all symmetries  of
the free field theory of $\Phi^i$ as is most easily seen from the unfolded
dynamics approach (see e.g. \cite{333}).

For an  algebra $A$ with the product law $\star$ (from now on by algebra we mean
associative algebra if not specified otherwise),  $l(A)$ denotes the associated Lie
(super)algebra with the (graded) commutator
\be
[a\,,b]_\star=a\star b\mp b\star a\quad\forall  a,b\in A
\ee
as a Lie product.
 Then the HS algebra $\hu(V_\Phi)$ is the real
form of $l(H(V_\Phi))$ singled out by the  conditions
\be
\label{sa}
\sigma (a) = a
\ee
with such conjugation $\sigma$ of $l(H(V_\Phi))$ ($\sigma(ia)=-i \sigma( a)$, $\sigma^2=Id$) that
the corresponding symmetry transformations of
$V_\Phi$ are unitary.\footnote{Note that the antihermiticity condition
implying unitarity of symmetry transformations  requires
the conjugation $\gs$ of $l(H_\Phi)$ be generated by
an involution $\dagger$ of $H_\Phi$ via $\gs (a):= -a^\dagger$
in the  Lie algebra (\ie bosonic) case. Since involution reverses the order of product factors,
$(ab)^\dagger= b^\dagger a^\dagger $, the antihermiticity condition
does not allow us to start with a real {} algebra $H_\Phi$
(anti-Hermitian operators do not form an associative algebra).}
Given conformal fields $\Phi$ in $d$ dimensions,
$\hu(V_\Phi)$ can be interpreted either as conformal HS algebra in $d$
dimensions or as $AdS_{d+1}$ HS algebra.

$hu(V_\Phi)$ admits further truncations
of the orthogonal  and symplectic types
induced by an involutive  antiautomorphism $\rho$ of $H(V_\Phi)$
\be
\label{rho}
\rho (ab) =\rho(b) \rho(a)\q \rho^2 = Id\,.
\ee
For $\Phi^i$ carrying color index $i=1,2,\ldots n$,  there are two options
for the extension  of $\rho$ that lead to two types
of HS algebras. Namely, if $\rho$ was an antiautomorphism
of the model with a single field $\Phi$, its color extension
$\rho^{col}$  is
\be
\label{rcol}
\rho^{col} (\Phi^i) = \eta_{ij} \rho(\Phi^j)\,,
\ee
where $\eta_{ij}$ is some nondegenerate matrix.
(Note that $\rho^{col}$ maps left and right
$H(V_\Phi)$-modules to each other.)
The truncation condition is
\be
\label{hou}
\rho^{col} (a) = - i^{p(a)} a\,,
\ee
where $p(a)= 0$ or $1$ is the boson-fermion parity of $a$.
Depending on whether $\eta_{ij}$ is symmetric or antisymmetric, this gives
the algebras
$\ho(V_\Phi)$ or $\husp(V_\Phi)$, respectively. In the case of a single field $\Phi$
$(n=1)$, $\ho(V_\Phi)$ is the minimal HS algebra.
 For  HS algebras associated with symmetric fields of integer spins, $\ho(V_\Phi)$
contains  even spins. Note that if $\eta_{ij}$
has no definite symmetry, the subalgebra singled out by
(\ref{hou}) is a direct sum of algebras $\ho(V_\Phi)$ and $\husp(V_\Phi)$
with smaller $n$. (For more detail we refer the reader to \cite{Ann,Konstein:ij}
where this construction was originally applied to $AdS_4$ HS algebras.)

HS algebras available in the literature belong to the  three classes
$\hu(n_{s_1},n_{s_2},\ldots |d )$,\\ $\ho(n_{s_1},n_{s_2},\ldots |d )$ and
$\husp(n_{s_1},n_{s_2},\ldots |d )$ with $0\leq s_1 <s_2 <s_3\ldots $.
Here $d$ is dimension of space-time where a set  of conformal fields $\Phi^i$
contains $n_{s_1}$ fields of spin $s_1$, $n_{s_2}$ fields of spin $s_2$, etc.
$\hu(n_{s_1},n_{s_2},\ldots |d )$ is  the algebra $hu(V_\Phi)$ for the corresponding set
of fields $\Phi^i$
while $\ho(n_{s_1},n_{s_2},\ldots |d )$  and $\husp(n_{s_1},n_{s_2},\ldots |d )$
are its subalgebras singled out by condition (\ref{rcol}). In principle, one can also consider the
case where  different fields $\Phi^i$ live in space-times of different dimensions.
Though no algebras of this type were so far considered in the literature, recent results of
\cite{Gelfond:2010pm}, where it was shown that current interactions of $4d$ massless fields
 acquire natural interpretation in terms of a mixed system of $4d$ and $6d$
 conformal fields, suggest that they may also be of interest.
 For such algebras one can use notation $h_{\cdots}(n_{d_1 s_1},n_{d_2 s_2},\ldots )$.

The list of ghost-free propagating conformal fields in $d$ dimensions depends
on whether $d$ is even or odd. As shown in \cite{Siegel,Mets},
apart from massless scalar and spinor in
any dimension, only mixed symmetry fields with field
strengths described by rectangular Young diagrams of height $d/2$
in even space--time dimension correspond to unitary theories.
Hence, for even $d$, conformal massless fields are characterized by a single
spin parameter $s$ associated with a length $s$ or $s-\half$  of tensor or
spinor-tensor Young diagram, respectively.
Conformal scalar and spinor correspond
to Young diagrams of zero length, hence  making sense for odd $d$ as well.

The presented construction of HS algebras is closely related to Flato-Fronsdal-type
theorems on the relation between tensor product of conformal fields in $d$
dimensions and massless fields in $AdS_{d+1}$. Indeed, HS gauge fields associated
with the $AdS_{d+1}$ HS algebra are valued in the algebra of operators that act in
$V_\Phi$,
which is $V_\Phi^*\otimes V_\Phi$ as a linear space. An important feature of HS gauge theories
is \cite{Ann,Vasiliev:2003ev} that Weyl 0-forms, which contain all degrees of freedom of the
$AdS_{d+1}$ system, are valued in the so-called twisted adjoint module which is isomorphic to
$V_\Phi^*\otimes V_\Phi$ as a linear space. This implies that degrees of freedom of the
$AdS_{d+1}$ HS theory belong to the module equivalent to the tensor square of
the conformal module $V_\Phi$ in $d$ dimensions (here we do
not distinguish between vector spaces $V^*_\Phi$ and $V_\Phi$ which are isomorphic
in the unitary case). Hence, the  construction of HS algebras
is such that the spectrum of fields of the bulk HS theory is designed to
result from the tensor product of boundary conformal fields.
This is  in kinematical agreement
with the idea of $AdS/CFT$ correspondence because  $V_\Phi\otimes V_\Phi$
is the space of  conformal conserved currents of the theory of free fields
 $\Phi^i$ as is most directly seen in the unfolded
dynamics approach  \cite{tens2}, which fact is of course not surprising given that
$V_\Phi\otimes V_\Phi$ is the space of conformal HS symmetries.
Hence  HS algebras $h_{\cdots}(n_{s_1},n_{s_2},\ldots |d)$
are properly designed to support $AdS/CFT$ correspondence
between boundary conformal theories and bulk HS theories, the issue  which acquired
a lot of interest during recent years. (See, e.g.,
\cite{Sundborg:2000wp,WJ,Konstein:2000bi,Vasiliev:2001zy,Mikhailov:2002bp,Sezgin:2002rt,Klebanov:2002ja,Sezgin:2003pt,
Giombi:2009wh,Metsaev:2008fs,Bekaert:2010ky,
Joung:2011xb,Giombi:2011kc,Aharony:2011jz,Henneaux:2010xg,Campoleoni:2010zq,Gaberdiel:2010pz,
Gaberdiel:2011wb,Ahn:2011pv,Gaberdiel:2011zw,Chang:2011mz,Gaberdiel:2011nt,Jevicki:2011ss,
Anninos:2011ui,Campoleoni:2011hg,
Kraus:2011ds,Ammon:2011ua,Maldacena:2011jn,Gaberdiel:2012yb,Vasiliev:2012vf,
Maldacena:2012sf,Gupta:2012he,Alkalaev:2012rg,Didenko:2012vh,Chang:2012kt,Colombo:2012jx,Didenko:2012tv}.
For reviews and more references see also \cite{Gaberdiel:2012uj,Giombi:2012ms,Jevicki:2012fh}.)
Of course,
being applicable to free fields, the above consideration has to be reanalyzed at the
interaction level. For example, as shown in \cite{Vasiliev:2012vf}, except for two particular reductions
of the HS gauge theory which, in accordance with the theorem of Maldacena and Zhiboedov
\cite{Maldacena:2011jn}, are dual to the   boundary theory of free currents, all other nonlinear
 $AdS_4$ HS gauge theories turn out to be dual to a $3d$ conformal HS gauge theory of interacting
 currents.

Although original $AdS_4$ HS algebras were obtained in
\cite{Fradkin:1986ka,Vasiliev:1986qx,Fradkin:1987ah} from  different arguments
\cite{Vasiliev:1986td} aimed at  reformulation of HS theory  in terms of
differential forms, that eventually led to its unfolded formulation \cite{Ann,con,more},
they belong to the class of HS algebras discussed above.
Even with no reference to symmetries of $3d$ unitary conformal fields,
original $AdS_4$ HS algebras were interpreted as  algebras of $3d$ conformal HS gauge theory
 by Fradkin and Linetsky in \cite{Fradkin:1989xt}.

Specifically, the original bosonic $AdS_4$ HS algebra found in \cite{Fradkin:1986ka}
is $\hu(1_{0} |3 )$. Its extension to $\hu(1_{0}, 1_{1/2} |3 )$ was proposed in
\cite{Fradkin:1987ah} and to $h_{\cdots} (n_{0}, m_{1/2} |3 )$ in \cite{Konstein:ij}
 where they were called $h_{\cdots}(n, m |4 )$.
Other way around, $4d$ conformal HS algebras introduced
by Fradkin and Linetsky in \cite{FL} in the context of
$4d$ conformal HS gauge theory were later interpreted  as  $AdS_5$ HS algebras in
 \cite{Sezgin:2001zs,Vasiliev:2001wa,Sezgin:2001yf,Alkalaev:2002rq}.
 All these algebras are $h_{\cdots} (n_{0}, n_{1/2},n_1 |4 )$ where  $n_0 ,n_{1/2}$
 and $n_1$ are numbers of fields of respective spins in a supermultiplet of
 $4d$ $N$-extended conformal superalgebra with $N=1,2,4$
including the case of $N=4$ SYM multiplet  $\hu(n_{0},n_{1/2},n_1 |4 )$
with $n_0=6n_1$, $n_{1/2}=4n_1$.
 Their realization in terms of  $4d$ boundary fields was considered in \cite{Vasiliev:2001zy}
 including  the generalization to $h_{\cdots}(n_{s_1},n_{s_2},\ldots |4 )$
 with nonzero $n_s$ at $s>1$. Extension  to $h_{\cdots}(n_{s}|d )$
 with any even $d\geq 4$ was given in \cite{Bekaert:2009fg}. Algebra
 $\ho(1_0|6)$ interpreted as the minimal $AdS_7$ HS algebra was considered in
 \cite{Sezgin:2001ij}. In \cite{Vasiliev:2001zy},  algebras
$h_{\cdots}(1_{0},1_{1},1_2\ldots \infty |4 )$ and
$h_{\cdots}(1_{0},1_{1/2},1_1\ldots\infty |4 )$ (the case of $M=4$ in notations of
\cite{Vasiliev:2001zy}) and  $h_{\cdots}(1_{0},1_{1},1_2\ldots \infty |6 )$ and
$h_{\cdots}(1_{0},1_{1/2},1_1\ldots\infty |6 )$ (the case of $M=8$ in notations of
\cite{Vasiliev:2001zy}) were identified with $l(A_M)$ where Weyl algebra $A_M$ is the
algebra of various polynomials of $M$ pairs  of oscillators.
 $\hu(1_{0} |d )$ was
identified as the algebra of conformal HS symmetries of a massless scalar by Eastwood
in \cite{Eastwood:2002su} and was used for the construction of HS gauge theories in $AdS_{d+1}$ in
\cite{Vasiliev:2003ev} where it was also extended to $\hu(n_{0} |d )$. HS superalgebras
$\hu(n_{0},m_{1/2} |d )$ were introduced in \cite{Vasiliev:2004cm}.

All HS algebras listed above admit  realizations in terms of
 Weyl algebras, which  are particularly useful for the
formulation of nonlinear HS gauge theories of \cite{more,Vasiliev:2003ev}.
There are two types of constructions
mentioned in Introduction. The one with elementary spinor oscillators was used in
\cite{Vasiliev:1986qx,Fradkin:1987ah,Fradkin:1989xt,FLA,
Konstein:ij,Sezgin:2001zs,Vasiliev:2001wa,Vasiliev:2001zy,Sezgin:2001yf,
Alkalaev:2002rq}. That with elementary oscillators carrying vector indices used in
 \cite{Vasiliev:2003ev,Vasiliev:2004cm,Bekaert:2009fg} applies to HS models
 in any dimension. Being closely related to twistor theory,
the spinor realization is likely to be both simpler and deeper.

For example, HS algebras $\hu(n_0,m_{1/2}|3)$ were shown in \cite{Konstein:ij}
(where they were denoted $hu(n,m|4)$)
to be realized by matrices
\begin{equation}
\label{P}
P_i{}^j({Y})={\kern 2em\raise
2.5ex\hbox{$n$}}{\kern -0.7em\raise -2.5ex\hbox{$m$}}\phantom{n}
\begin{tabular}{|l|l|} \multicolumn{2}{l}{\kern 2.5em\hbox{$n$}\kern
6em\hbox{$m$}} \\ \hline $P^E{}_{i^\prime}{}^{j^\prime}
(Y)\phantom{\biggl(}$  &
$P^O{}_{i^\prime}{}^{j^{\prime\prime}}(Y)$ \\
\hline
$P^O{}_{i^{\prime\prime}}{}^{j^\prime}(Y)
\phantom{\biggl(}$
& $P^E{}_{i^{\prime\prime}}{}^{j^{\prime\prime}}(Y)$ \\
\hline
\multicolumn{2}{c}{}
\end{tabular}
\end{equation}
where matrix valued polynomials $P^E(Y)$ and  $P^O(Y)$
are respectively even and odd functions of the oscillators $Y_A$
($A=1,2,3,4$ is the $4d$ Majorana spinor index) that obey the star-product
commutation relations (\ref{weyl1})
where $C_{AB}$ is the $4d$  charge conjugation matrix.

The space $V_\Phi$
of single-particle states of $n$ massless scalars and $m$ massless spinors
in three dimensions is realized as the direct sum of $n$ even subspaces $F_0$
and $m$ odd subspaces $F_1$ of the Fock module $F$
\be
F:\quad f(Y^a_+) |0\rangle\qquad Y_-^a |0\rangle =0\,,
\ee
where $Y^a_\pm$ is a pair of
mutually conjugated canonical oscillators in the set $Y^A$. $F_0$ and $F_1$ are
 spanned, respectively, by even and odd functions $f(Y^a_+)$.

That HS algebras are naturally realized in terms of  Weyl algebra is not accidental. As
algebra of endomorphisms of a single-particle
space $V_\Phi$ of conformal fields, the HS algebra can be represented by differential
operators of various degrees acting in $V_\Phi$. The latter belong to the Weyl algebra
which is the algebra of various differential operators with polynomial coefficients.
Then $V_\Phi$ is represented as a Fock module of the star-product algebra or some its
quotient.

The list of full nonlinear HS gauge theories with propagating HS gauge fields known
so far, that admit HS algebras as algebras of symmetries of their maximally symmetric
vacua, is much shorter than the list of HS algebras given above.
It includes  $AdS_4$ theories based on
$h_{\cdots} (n_0,m_{1/2}|3)$ \cite{con,more} and
$AdS_{d+1}$ theories based on $h_{\cdots} (n_0|d+1)$ \cite{Vasiliev:2003ev}.
(For  HS theories in $AdS_{d+1}$ with $d\leq 2$ not considered in this paper,
 where HS gauge fields carry no degrees of freedom, see \cite{Vasiliev:1999ba}
 and references therein.)
The problem of construction of full
nonlinear HS theories associated with other HS algebras remains open, though
some partial results on the construction of cubic interactions in the respective theories
were obtained in \cite{Vasiliev:2001wa,Alkalaev:2002rq,Alkalaev:2010af}.

Although the construction of HS algebras sketched above can be used for
description of particular mixed symmetry fields
(see also \cite{Boulanger:2011se}), it can unlikely be applied to
the generic case rich enough to incorporate String Theory. Hence,
some strategy change is needed. Before going into technical detail
in the next section, we comment on the general idea.

In the literature (see, e.g., \cite{Boulanger:2008up,Boulanger:2011se,Bekaert:2011js}),
 the construction of HS algebras is often related to
the universal enveloping algebra $U(s)$ of the space-time (conformal) symmetry
(super)algebra $s$. If the
 $s$-module $V_\Phi$ is irreducible,  $H(V_\Phi)$ is isomorphic to
$U(s)/I_{V_\Phi}$ where $I_{V_\Phi}$ is the ideal of $U(s)$ which consists of
those its elements that annihilate $V_\Phi$. This is tautologically the case
for a single field $\Phi$ since this just means that its single-particle
states form an irreducible $s$-module.
However,  identification of HS algebras with quotients of $U(s)$
may  be misleading  because the algebras
$H(V_\Phi)$ differ from $U(s)/I_{V_\Phi}$ for reducible  ${V_\Phi}$.
Indeed, since $H(V_\Phi)$ is the maximal algebra acting on ${V_\Phi}$,
\be
U(s)/I_{V_\Phi}\subset H(V_\Phi)\,.
\ee
Isomorphism $H(n_0, n_{1/2},\ldots )\sim U(s)/I_{V_\Phi}$
takes place only for irreducible $V_\Phi$, \ie
 iff $n_{s_0}=1$ and $n_{s}=0$ at $s\neq s_0$ for some $s_0$.

Given {} algebra $A$,
we introduce associative  multiparticle algebra $M(A)$ as $U(l(A))$.
The {} algebra $H(V_\Phi)$ of endomorphisms of $V_\Phi$,
which underlies the construction of HS algebra $\hu(V_\Phi)$,
gives rise to $M(H(V_\Phi))$. Being defined as universal enveloping of
 $\hu(V_\Phi)$, $M(H(V_\Phi))$ acts on every $\hu(V_\Phi)$-module.
In particular, it acts on the space
\be
\label{tensvphi}
\cV_\Phi = \sum_{n=0}^\infty \oplus V_\Phi^n\q V_\Phi^n=
Sym \,  \underbrace{V_\Phi\otimes \ldots \otimes V_\Phi}_n\,
\ee
which is nothing else but the space of all multiparticle states of the fields $\Phi$.
Multiparticle algebra associated with the fields $\Phi$
 will be identified with the appropriate real form of the Lie (super)algebra
 $l(M(H(V_\Phi)))$ or some of its quotients  considered in Section \ref{genw}.

 As discussed in Section \ref{modul},
 apart from the simplest possibility where $l(M(H(V_\Phi)))$ acts independently
 on every $V_\Phi^n$, multipartcle algebras admit representations mixing
 $V_\Phi^n$ with
 different $n$. Relating multiparticle states of the field  theory of $\Phi$
 such as, e.g., $N=4$ SYM theory, multipartcle algebras look particularly
 appealing in the String Theory  context. The problem of  increase of lowest
 energies discussed in Introduction is avoided because a multiparticle
 algebra contains $\hu(V_\Phi)$ as  subalgebra that acts on $V_\Phi\in \V_\Phi$
 as in the original
 HS theory, hence having the same weights (in particular, energies) in this sector.
The same time, that
 the multiparticle algebra  is much smaller than the maximal symmetry algebra
 $\hu(\V_\Phi)$ acting on the space of all multiparticle states of $\Phi$  should
 leave enough flexibility for description of interacting fields $\Phi$.

\section{Associative multiparticle algebra}
\label{genw}
\subsection{Definition}
\label{setup}
Let $A$ be an algebra with the product law $\star$ and basis elements $t_i$ obeying
\be
\label{struc}
t_i \star t_j = f^k_{ij} t_k\,.
\ee
 Associativity  implies
\be
\label{ass1}
f_{ij}^k f_{kl}^n = f_{ik}^n f_{jl}^k\,.
\ee
In  HS algebras, $t_i$ denotes the infinite set of elements
 $t_i=(1, Y^A, Y^A Y^B, \ldots )$ while
$\star$ is the star product on functions of $Y$.

Algebra $M(A)$ is defined as follows.
As a linear space, it is isomorphic to  direct sum
of all symmetric tensor degrees of $A$
\be
\label{tens}
M(A) = \sum_{n=0}^\infty \oplus Sym \,  \underbrace{A\otimes \ldots \otimes A}_n\,.
\ee
A natural basis of $M(A)$ is provided by  symmetrized tensor product monomials
\be
\label{tenst}
T_{i_1\ldots i_n} =  Sym \,t_{i_1}\otimes \ldots \otimes t_{i_n}\q
T_{\ldots j\ldots k\ldots  }=T_{\ldots k\ldots j\ldots  }\quad \forall j,k\,.
\ee

Let $A^*$ be the space of linear functionals on $A$, \ie
\be
\ga\in A^* :\quad \ga=\sum_i \ga_i t^{*i}\,,\quad t^{*i} (t_j) = \delta ^i_j\,,
\ee
where $\{t^{*i}\}$ is the basis of $A^*$ dual to $\{t_i\}$.
$M(A)$ is the algebra of functions $F(\ga)$ on $A^*$
with the product law
\be
\label{circ}
F(\ga)\circ G(\ga) = F(\ga) \exp \Big (  \f{\overleftarrow{\p}}{\p \ga_i} f^n_{ij} \ga_n
\f{\overrightarrow{\p}}{\p \ga_j} \Big ) G(\ga)\,,
\ee
where derivatives $\f{\overleftarrow{\p}}{\p \ga_i}$ and $\f{\overrightarrow{\p}}{\p \ga_j}$
act on $F$ and $G$, respectively.
An elementary computation gives
\be
((F_1\circ F_2 )\circ F_3)(\ga) =\exp \Big (f^n_{ij}\ga_n \sum_{\gamma< \gb =1,2,3}
  \f{\p^2}{\p \ga_{\gamma i}\p \ga_{\gb j}} +f^n_{ij}f^k_{nm}\ga_k \f{\p^3}{\p \ga_{1i}\p \ga_{2j}
  \p \ga_{3 m}}\Big ) F_1 (\ga) F_2(\ga)  F_3(\ga)\,,
\ee
\be
(F_1\circ (F_2 \circ F_3))(\ga) =\exp \Big (f^n_{ij}\ga_n \sum_{\gamma< \gb =1,2,3}
  \f{\p^2}{\p \ga_{\gamma i}\p \ga_{\gb j}} +f^k_{in}f^n_{jm}\ga_k \f{\p^3}{\p \ga_{1i}\p \ga_{2j}
  \p \ga_{3 m}}\Big )
F_1 (\ga) F_2(\ga)  F_3(\ga)\,,
\ee
where $\f{\p}{\p \ga_{\gb i}}$ acts on $F_\gb (\ga)$ $(\gb=1,2,3)$. Hence, the
$A$-associativity (\ref{ass1}) implies associativity of the product
$\circ$ of $M(A)$.\footnote{Note that somewhat similar  (though different in
some important details) algebras of oscillators  were mentioned in
\cite{Engquist:2005yt} in the context of singleton strings aimed at
the description of multisingleton states which is another name for
 multiparticle states.} Note that $A\subset M(A)$ is represented by
linear functions on $A^*$. Hence,  $\star$  product acts on linear
functions on $A^*$ according to
\be\label{staralfa}
\ga_j \star\ga_k  =  f_{jk}^m\ga_m \,.
\ee
Note however that $A\circ A$ does not belong to $ A$.

Algebra $M(A)$ is unital, with the unit element $Id$ identified with $F(\ga)=1$. Hence
\be
\label{mpr}
M(A) =\mathbb{K} \subset{\ls\, \mbox{\small +}}  M^\prime(A) \,,
\ee
where $\mathbb{K}$ is the field over which $A$ and $M(A)$ were defined.
(In  HS context, the most important case is $\mathbb{K}=\mathbb{C}$.)
Indeed, from (\ref{circ}) it follows that unit element never
appears on the \rhs of $F\circ G$ if  $F(\ga)$ and/or $G(\ga)$ is
a homogeneous monomial of non-zero degree.

The $\mathbb{Z}_2$ grading of $M(A)$ is induced by that of $A$
\be
F((-1)^{\pi (\ga)}\ga)= (-1)^{\pi(F)} F(\ga)\,.
\ee

 $M(A)$ is isomorphic to the universal enveloping algebra $U(l(A))$
\be
\label{iso}
M(A)\sim U(l(A))\,.
\ee
Indeed, by definition of  $l(A)$,
\be
[t_i\,,t_j]_\star = g_{ij}^k t_k\q
g_{ij}^k = f_{ij}^k- f_{ji}^k\,.
\ee
On the other hand, from  (\ref{circ}) it follows that
\be
\label{tcirct}
\ga_i\circ \ga_j - \ga_j\circ \ga_i = g_{ij}^k \ga_k\,.
\ee
Along with associativity of $M(A)$,  Eq.~(\ref{mpr}) and the fact that $M(A)$
is isomorphic to $U(l(A))$ as a linear space,
Eq.~(\ref{tcirct}) proves  (\ref{iso}).
Concise form of the product law (\ref{circ}) is specific to
the case where a Lie algebra  $l$ of $U(l)$ is associated with an
associative algebra $A$, \ie $l=l(A)$.

 The following useful property of $M(A)$  is a simple
consequence of Eq.~(\ref{circ})
\be
\label{fr} \forall f,g\in
A:\qquad \exp f(\ga) \circ \exp g(\ga) =
\exp(f\bullet g)(\ga)\,,
\ee
where
\be \label{bull} f\bullet g := f+g +f\star g =
(f+e_\star)\star (g+e_\star)-e_\star\in A\,
\ee
and $e_\star$ denotes the unit element of $A$ if the latter is unital
 (recall that $f,g\in A$ implies that $f(\ga)$ and $g(\ga)$ are linear
in $\ga$).
Associativity of $\star$ implies associativity of $\bullet$
\be
(f\bullet g)\bullet h = f\bullet (g\bullet h) = (f+e_\star)\star (g+e_\star)\star (h+e_\star)-e_\star\,.
\ee
Note that the product $\bullet$ is associative even if $A$ is not unital.

Let
\be
\label{fnu}
G_\nu =\exp(\nu)\in M(A)\q \nu=\nu^i\ga_i\,,
\ee
where $\nu^i\in \mathbb{K}$ are free parameters.
Eq.(\ref{fr}) gives
\be
\label{fnumu}
G_\nu\circ G_\mu = G_{\nu\bullet \mu}\,.
\ee
This formula is convenient for practical computations with $G_\nu$ used as
the generating function for elements of $M(A)$ resulting from differentiation
over $\nu^i$.

\subsection{Linear maps}
\label{Linear maps}
Algebra $M(A)$ is double filtered in the following sense.
Let $V_n$ be the linear space of order $n$  polynomials of $\ga_i$.
{}From Eqs.~(\ref{circ}), (\ref{tcirct}) it follows that for any $F_n\in V_n$
and $F_m\in V_m$
\be
F_n \circ F_m \in V_{n+m}\q F_n \circ F_m -F_m \circ F_n \in V_{n+m-1}\,.
\ee
This property holds for any universal enveloping algebra (see, e.g.,
\cite{diks}).

A  linear map of $M(A)$ to itself is represented by
\be
\U(\ga, a)=\sum_{m,n=0}^\infty \U^{i_1\ldots i_m}{}_{j_1\ldots j_n}\ga_{i_1}\ldots
\ga_{i_m} a^{j_1}\ldots a^{j_n}\,
\ee
with
\be
\U(\ga, a) [F]= \U(\ga, \f{\overrightarrow{\p}}{\p \ga}) F(\ga)\,,
\ee
where derivatives $\f{\overrightarrow{\p}}{\p \ga}$ act on $F(\ga)$.
This formula can be interpreted as representing the action of the normal-ordered
oscillator algebra with the generating elements $\ga_i$ and $a^j$
acting on the Fock module spanned by $F(\ga)|0\rangle$ with
$a^i|0\rangle =0$.
To respect the double filtration property, mapping order-$n$ polynomials to
order-$n$ polynomials, $\U(\ga, a)$ should obey
\be
\U^{i_1\ldots i_m}{}_{j_1\ldots j_n}=0 \quad \mbox{at}\quad m>n\,.
\ee
Maps of this class, which we call filtered, are of most interest
in this paper.

In these terms, the unit map is
$
\mathbf{Id} = 1\,.
$
The map induced by a linear map $u(t_i)=u_i{}^j t_j$ of $A$  is represented by
\be
\label{u}
\U(\ga, a) = \exp(\ga_i u_j{}^i a^j-\ga_i a^i)\,.
\ee

Consider maps of the form
\be
\label{Uf}
U(f) \equiv \U(\ga,a|f)= \phi \exp (\ga_i f^i(a))
\ee
with some $\ga$-independent coefficients $f^i(a)$ and constant $\phi$.
The map $U(f)$ is filtered provided that $f^i(a)$ is at least linear in
$a$, \ie
\be
\label{atl}
f^i(0) =0\,.
\ee

Interpreting
$a$ as parameters, we can identify any $f(\ga)=\sum_i f^i \ga_i \in M(A)$
with $f(t)=\sum_i f^i t_i \in A$. For $U(f)$ (\ref{Uf}) acting on $G_\nu$ (\ref{fnu})
we obtain
\be
\label{uf}
U(f)(G_\nu) =\exp(\tilde f^i(\nu)\ga_i)\q \ \tilde f^i(\nu)=
\nu^i +f^i(\nu)\,.
\ee
Hence, Eq.~(\ref{fr}) gives
\be
\label{ufuf}
U(f)(G_\nu)\circ U(g)(G_\mu) =
\exp((\tilde f(\nu)\bullet \tilde g(\mu))(\ga))\,,
\ee
where $\tilde f(\nu)$ and $\tilde g(\mu)$ are now interpreted as elements of $A$,
\ie
$
\tilde f(\nu) =
\tilde f^i(\nu)\ga_i\,.
$

Important classes of linear maps $U$ of $M(A)$ onto
itself are represented by automorphisms
\be
\mathcal{T}(G_1\circ G_2) = \mathcal{T}(G_1)\circ  \mathcal{T}(G_2)
\ee
and antiautomorphisms
\be
\mathcal{R}(G_1\circ G_2) = \mathcal{R}(G_2)\circ  \mathcal{R}(G_1)\,
\ee
for $\forall G_{1,2}\in M(A)$.
To see whether or not $\mathcal{T}$ and $\mathcal{R}$ are, respectively,
automorphism and antiautomorphism of $M(A)$, it is enough to check
these properties for
$G_1=G_\nu$ and $G_2=G_\mu$ with arbitrary $\nu$ and $\mu$, hence
 solving the equations
\be
\mathcal{T}(G_\nu)\circ \mathcal{T}(G_\mu ) =
\mathcal{T}(G_{\nu\bullet \mu} )\,,
\ee
\be
\label{anti}
\mathcal{R}(G_\mu)\circ \mathcal{R}(G_\nu ) =
\mathcal{R}(G_{\nu\bullet \mu} )\,.
\ee

Let $\tau$ and $\rho$ be, respectively, an automorphism
and antiautomorphism of $A$, \ie
\be
\tau(a\star b) = \tau (a)\star \tau(b)\q
\rho(a\star b) = \rho(b)\star \rho(a)\qquad \forall a,b \in A\,.
\ee
In terms of basis elements $t_i$ and structure coefficients $f_{ij}^k$ this means
that matrices $\tau_i{}^j$ and $\rho_i{}^j$ defined via
\be
\tau(t_i) = \tau_i{}^j t_j\q \rho(t_i)=\rho_i{}^j t_j
\ee
obey
\be
\tau_i{}^{i^\prime } \tau_j{}^{j^\prime } f^k_{i^\prime j^\prime}= f^{k^\prime }_{ij}
\tau_{k^\prime }{}^k\q
\rho_i{}^{i^\prime } \rho_j{}^{j^\prime } f^k_{i^\prime j^\prime}= f^{k^\prime }_{ji}
\rho_{k^\prime }{}^k \,.
\ee
Eq.~(\ref{circ}) implies that $\tau$ and $\rho$ induce automorphism
${\mathcal T}$ and antiautomorphism ${\mathcal R}$ of $M(A)$
\be
\label{TR}
{\mathcal T} (F(\ga)) = F(\tau(\ga))\q
{\mathcal R} (F(\ga)) = F(\rho(\ga))\,.
\ee
These maps are described by $U(a,\ga)$ (\ref{u}) with $u_i{}^j$ identified
 either with $\tau_i{}^j$ or with $\rho_i{}^j$.

Analogously, one proceeds for conjugation $\sigma$ and involution $\dagger$
which are antilinear (\ie conjugating complex numbers)
counterparts of automorphism and antiautomorphism, respectively,
\be
\label{sdag}
{\mathcal S} (F(\ga)) = \bar F(\sigma(\ga))\q
 (F(\ga))^{\ddag} = \bar F(\ga^\dagger)\,,
\ee
where $\bar F$ is complex conjugated to $F$, \ie the coefficients of the expansion
in powers of $\sigma (\ga )$ and $\ga^\dagger$ are complex conjugated to those of
the expansion in powers of $\ga$.

Consider maps (\ref{Uf}) with
\be
f^i(a)t_i = f(a_n)\q a_{n} =
\underbrace{a\star \ldots\star a}_n\q a_{1}=a=t_i a^i\q a_0=e_\star  \,,
\ee
where $f(a_n)$ is a linear function of $a_n$ ($n\geq 1$).
Such maps have the form (\ref{Uf}) since $a_n\in A$.

A particularly important subclass of maps (\ref{Uf}) is represented by
$\mathbf{U}_{u}$ of the form
\be
\label{ubgb}
\mathbf{U}_{u}(a) = \exp[ u(a) -a]\,,
\ee
$a\in A$ and
\be
\label{u(f}
u(a)=
(u_1{}^1 a+u_1{}^2 e_\star )\star
(u_2{}^1 a+ u_2{}^2 e_\star)_\star^{-1}\q (e_\star +\gb a)_\star^{-1} :=
\sum_{n=0}^\infty (-\gb
)^n a_n \,
\ee
with $u_i{}^j\in \mathbb{K}$. Composition of such maps gives  a map of the
same class
\be
\mathbf{U}_{u}\mathbf{U}_{v}= \mathbf{U}_{uv}\,,
\ee
where $(uv)_i{}^j= u_i{}^k v_k{}^j$ is the matrix product in $Mat_2(\mathbb{K})$.
The maps $\mathbf{U}_{u}$ with $det |u|\neq 0$ are invertible and form
 usual Mobius group.

{}From Eq.~(\ref{uf}) it follows that
\be \label{munu} \mathbf{U}_{u}
(G_\nu) = G_{{u}(\nu)}\,.
\ee
Consider the composition law of $M(A)$ in the basis
associated with
$
G_{{u}(\nu)},
$
assuming that new  basis elements, that replace (\ref{tenst}), are
\be
\label{bas}
T_{i_1\ldots i_n}^{u} = \f{\p^n}{\p \nu^{i_1}\ldots \p \nu^{i_n}}
G_{{u}(\nu)}\Big |_{\nu=0}\,.
\ee
To this end, we have to compute
\be
G_\nu \diamond G_\mu =
\mathbf{U}^{-1}_{u} ( G_{{u}(\nu)} \circ G_{{u}(\mu)})\,.
\ee
Eq.~(\ref{bull}) gives
\be
\label{diam}
G_{\nu} \diamond G_{\mu} =G_{u^{-1}(u(\nu)\bullet u(\mu))} \,.
\ee

Generally, maps (\ref{ubgb}) are not filtered, not respecting the condition
(\ref{atl}). The subgroup $P$ of filtered maps (\ref{ubgb}) is represented by
lower triangular matrices
\be
\label{aff}
u_{b,\gb}(f) = b f\star (e_\star+\gb f)^{-1}_\star\,
\ee
with the composition law
\be \label{grlow} b_{1,2}=b_1 b_2\q \gb_{1,2} = \gb_2 +\gb_1 b_2\,.
\ee
Clearly, $P$ is isomorphic to the affine group of translations and
dilatations of $\mathbb{R}^1$.

For affine transformations (\ref{aff}) we will  use notation
$U_{b,\gb}$ instead of $U_u$. In these terms, the unit element is
\be
\mathbf{Id} =\mathbf{U}_{1,0}\,
\ee
and
\be
\label{inv}
\mathbf{U}^{-1}_{b,\gb}= \mathbf{U}_{b^{-1}, -\gb b^{-1}}\,.
\ee

The map
\be
\label{RU}
{\mathbf{R}} =\mathbf{U}_{-1,1}\,
\ee
 is involutive
\be
\label{mathr}
\mathbf{R}^2 = \mathbf{Id}\,
\ee
and describes an antiautomorphism of $M(A)$. Indeed,
one can check (\ref{anti}) using that
\be
\mathbf{R}(G_\nu)\circ \mathbf{R}(G_\mu) =
\exp[(e_\star+\nu)^{-1}_\star-e_\star)\bullet ((e_\star+\mu)^{-1}_\star-e_\star)]
\ee
and, by (\ref{fnumu}),
\be
\mathbf{R}(G_{\mu}\circ G_{\nu})=
\exp ((e_\star+\mu\bullet \nu)^{-1}_\star-e_\star)\,.
\ee
{}Eq.~(\ref{munu}) gives
\be
\label{RF}
\mathbf{R} (G_\nu) = G_{- \nu\star (e_\star+ \nu)^{-1}_\star}\,.
\ee
Differentiation over $\nu^i$ gives, in particular,
\be
\mathbf{R}(Id) = Id\,,
\ee
\be
\label{R1}
\mathbf{R}(\ga_i) = - \ga_i\,,
\ee
\be
\label{R2}
 \mathbf{R}(\ga_i\ga_j) =\ga_i\ga_j + \{\ga_i\,,\ga_j\}_\star\,,
\ee
where, for simplicity, we consider the  even case with $\pi(\ga_i )=0$.

The antiautomorphism $\mathbf{R}$ of $M(A)$ exists independently of the specific
structure of $A$ and is called principal antiautomorpism of $U(l(A))$
\cite{diks}. Note that the form of $\mathbf{R}$ (\ref{ubgb}),
(\ref{RU}) is specific for  universal enveloping algebra of
a Lie algebra associated with an algebra $A$.

Given  algebra $A$, the {\it opposite} {}
algebra $\tilde A$ is isomorphic to $A$ as a linear space and has
the   product law $\tilde {\circ}$
\be
\label{tcirc}
a\,\tilde{\circ}\, b =  b\circ a\,.
\ee
An antiautomorphism $\rho$ of $A$ can be interpreted as the homomorphism
between $A$ and $\tilde A$. If $\rho$ is invertible, $\tilde A$ is
isomorphic to $A$. Hence, Eq.~(\ref{mathr}) proves
\be
 M(  \widetilde A) = \widetilde M( A) \sim M(A)\q \forall A\,,
\ee
which is of course in agreement with the realization of $M(A)$ as $U(l(A))$ \cite{diks}.

For affine maps, the composition law (\ref{diam}) takes the form
\be
\label{diamaf}
G_{\nu} \diamond G_{\mu} =G_{\gs_{b,\gb}(\nu,\mu)} \,,
\ee
where
\be
\label{sig}
\gs_{b,\gb}(\nu,\mu) = -\gb^{-1} (e_\star-(e_\star+\gb \mu )\star (e_\star-\gb (b+\beta) \nu\star \mu )^{-1}\star
(e_\star+\gb \nu ))\,.
\ee
In particular, this formula gives
\be
\gs_{1,0}(\nu,\mu) = \nu+\mu +\nu\star \mu=\nu\bullet \mu\,,
\ee
\be
\gs_{-1,1}(\nu,\mu) = \nu+\mu +\mu\star \nu= \mu\bullet \nu\,,
\ee
\be
\label{fus}
\gs_{1,-\half}(\nu,\mu) = 2(e_\star-(2e_\star -\mu)\star ( 4e_\star+ \nu\star \mu )_\star^{-1}
\star (2e_\star-\nu))\,.
\ee
Here $\gs_{1,0}(\nu,\mu)$ corresponds to the unit map, $\gs_{-1,1}(\nu,\mu)$
corresponds to the  antiautomorphism $\mathbf{R}$, while $\gs_{1,-\half}(\nu,\mu)$
describes the map reproducing the ${\mathcal N}\to 0$
limit of the $F$-current operator algebra of \cite{Gelfond:2013xt}.

\subsection{ Supertrace and central charge}
\label{auto}

Let $A$  possess a (super)trace $tr$ obeying
 \be
 \label{tr}
 tr(a\star b)=(-1)^{\pi(a)\pi(b)} tr(b\star a)\q \forall a, b\in A
\ee
($\pi(a)=0$ or $1$ is the $\mathbb{Z}_2$ grading of $a$; usual trace
is a particular case with $\pi(a)\equiv 0$.)
Let $A$ admit such a basis $t_i$  that
\be tr(\sum_i a^i t_i) = a^0\,,
\ee
\ie $tr (t_i) =
\delta_i^0$. Then
 \be g_{ij}= f^0_{ij}\, \ee
 is (graded)symmetric
\be g_{ij} = (-1)^{\pi_i\pi_j} g_{ji}\,. \ee Note
that $tr$ is supposed to be even, \ie $t_0$ is even which implies
that $g_{ij}$ is nonzero if $\pi_i = \pi_j$. If the bilinear
form $tr(a\star b)$ is non-degenerate, which is necessarily true if
$A$ is simple since zeros of $tr(a\star b)$ form a two-sided ideal
of $A$, $g_{ij}$ can be interpreted as a non-degenerate metric.
Associativity of $A$ implies via  $tr((t_i\star t_j)\star t_k)=
(-1)^{\pi_i} tr((t_j\star
t_k )\star t_i))$
graded cyclicity of the structure coefficients
 \be \label{cyc}
f_{ijk} = (-1)^{\pi_i} f_{jki}\q  f_{ijk}= f_{ij}^n
g_{nk}\,.
\ee

For unital algebra $A$, it is convenient to set $t_0=e_\star$ that is
reachable via rescaling of $tr$ in the non-degenerate case with
$tr (e_\star)\neq 0$. In the degenerate case with $tr(e_\star)=0$ a basis
element supporting trace differs from $e_\star$ analogously to the case
of $psu(2,2|4)$ familiar from $N=4$ SYM.  In that case,
$l(A)$ acquires an  ideal associated with $tr$ in addition to that
associated with $e_\star$. Note that, being related to $N=4$ SYM, the degenerate
case may  be of primary importance in the multiparticle extension
of HS theory.

Trace of $A$ induces a family of traces of $M(A)$. Indeed,
for exponentials $G_\nu $ (\ref{fnu}) define trace as
\be
\label{gtr}
Tr_{\Phi,\phi_\star} ( G_\nu ) = \Phi (tr (\phi_\star (\nu)))\q \phi_\star (\nu) =
\sum_{n=0}^\infty \phi_n \underbrace{\nu \star \ldots \star \nu}_n \,,
\ee
with any star-product function $\phi_\star (\nu)$ and usual function $\Phi(x)$.
{}From (\ref{fnumu}) it follows that
\be
Tr_{\Phi,\phi_\star} ( G_\nu\circ G_\mu) = \Phi(tr (\phi_\star (\nu\bullet \mu)))\,.
\ee
Using (\ref{tr}) and (\ref{bull}) it is easy to see that
$tr (\phi_\star (\nu\bullet \mu))=tr (\phi_\star (\mu\bullet \nu))$
and, hence,
\be
Tr_{\Phi,\phi_\star} ( G_\nu\circ G_\mu) =Tr_{\Phi,\phi_\star} ( G_\mu\circ G_\nu)\,.
\ee
Since $G_\nu$ is the generating function for any element of $M(A)$,
formula (\ref{gtr}) defines a trace of $M(A)$. Thus the space of traces
of $M(A)$ admits at least a freedom in two functions of one variable.
Note that the freedom in the definition of trace in $M(A)$
reflects the fact that $M(A)$ is not simple as is the case for every
universal enveloping algebra.

Remarkably, the freedom of the definition of $Tr$ of $M(A)$ is closely
related to the freedom in the central charge of the current operator
algebra. To reproduce the dependence on the
central charge the basis  has to be modified   further
by virtue of a field redefinition of the form
\be
\label{eta}
G_\nu \to  \eta(\nu) G_\nu\,,
\ee
where $\eta(\nu)$ is some map from $A$ to $\mathbb{K}$.
This map modifies the product law (\ref{diamaf}) to
\be
\label{diamaftrc}
\tilde G_{\nu} \diamond \tilde G_{\mu} =
\f{\tilde\eta(\mu)\tilde\eta(\nu)}{\tilde\eta(\gs_{b,\gb}(\nu,\mu))}
\tilde G_{\gs_{b,\gb}(\nu,\mu)}\q \tilde \eta(\nu) =\eta(u_{b,\gb}(\nu)) \,.
\ee
The form of the current OPE gets modified since the basis is still defined by
the formula analogous to (\ref{bas}) with respect to $\tilde G_\nu $.

As shown in Section \ref{Current operator algebra},
in the case of $F$-current algebra, the map (\ref{eta}) is defined so that
$Tr_{\Phi,\phi_\star} ( \tilde G_\nu)=1$ for certain $\Phi(x)$ and $\phi_\star$.
In the case of $A$-current algebra, the appropriate field redefinition is
still of the form (\ref{eta}), but it is not directly related with the
rescaling of some trace of $M(A)$.

\subsection{Ideals and quotients}
\label{ideals}
\subsubsection{Ideals induced by (anti)automorphism}
\label{quan}
Let $\tau$ be an automorphism of  $A$. A set of
elements that obey
\be
\tau(a)=a
\ee
forms a subalgebra $A_\tau$ of $A$. Suppose that $\tau$ is involutive,
\ie $\tau^2=Id$. Then $A_\tau$ consists of $\tau$-even elements
\be
a=\half (a+\tau(a))\,.
\ee
Let some $a\in A_\tau$ have the form
\be\label{el}
a=(b-\tau(b))\star c\q a\in A_\tau\q c\in A\,.
\ee
That $a\in A_\tau$ implies
\be
\label{pipi}
a=\half (b-\tau(b))\star (c-\tau (c))\,.
\ee
Elements (\ref{el}) form a two-sided ideal $I_\tau$ of $ A_\tau$.
Indeed,
\be
a\star (b-\tau(b)) = a\star b -\tau (a\star b) \quad \forall a\in A_\tau\,.
\ee
{}Eq.~(\ref{pipi}) implies that elements $a= b\star (c-\tau(c))$
form the same ideal $I_\tau$ of $A_\tau$. The algebra
\be
A^\tau = A_\tau / I_\tau\,
\ee
is spanned by those elements of $A_\tau$, that cannot be represented as
a product of $\tau$-odd elements of $A$.

In many cases, including usual HS algebras with  nontrivial
$\tau$, the latter condition turns out to be
too strong implying  $A^\tau =0$. For example, this is true for $A$ generated by
oscillators $Y^A$  treated as odd elements of the automorphism $\tau$
(which is the boson-fermion automorphism for spinorial $Y^A$). However, in the case of $M(A)$,
this construction leads to nontrivial result.

Let $\rho$ be an involutive antiautomorphism of $A$. As explained in Section \ref{HSA}, depending
on a particular choice of $\rho$, the condition (\ref{hou})  singles out the subalgebras
$\ho(V)$ or $\husp(V)$ of the HS
Lie algebra $\hu(V)$. In the general case let us call them $l_\rho(V)$.
Let $\mathcal{R}$ be the
antiautomorphism of $M(A)$  associated with $\rho$ via (\ref{TR}). Then
\be
\label{TRR}
\mathbf{T} = \mathbf{R}\mathcal{R}\,,
\ee
where $\mathbf{R}$ is the principal antiautomorphism (\ref{RU}),
is an involutive automorphism of $M(A)$.
It is not hard to see that $M^\mathbf{T}(A)\sim U(l_\rho(V))$.
Indeed, using (\ref{R1}),
the condition $\mathbf{T}(\ga_i) = \ga_i$ implies in particular
\be
\label{RA}
\rho(\ga_i) = - \ga_i \,,
\ee
which is just Eq.~(\ref{hou}).
As a result, all $\rho$-even elements of $A$ do not belong to $M^\mathbf{T}(A)$.
Factorization of the ideal $\mathbf{I}_\mathbf{T}(A)$ takes away the dependence on all
$\rho$-even elements of $A$.  As the algebra of functions
of $\rho$-odd elements of $A$, $M^\mathbf{T}(A)\sim U(l_\rho(A))$.

Using approach of Section \ref{Linear maps}, it is not difficult to
obtain explicit formulae for the composition law of $M^\mathbf{T}(A)$ in the form
analogous to (\ref{diamaf}). Indeed, consider basis (\ref{munu}) associated with
$G_{u_{1,-1/2}(\nu)}$. Impose the condition
\be
\label{rhod}
 \rho(\nu)=-\nu\,,
\ee
which is nothing else but
 the factorization condition removing dependence on $\rho$-even elements.
It is not difficult (but fun) to see that so defined elements $G_{u_{1,-1/2}(\nu)}$
are $\mathbf{T}$ invariant, \ie
\be
\mathbf{T} G_{u_{1,-1/2}(\nu)}=G_{u_{1,-1/2}(\nu)}
\ee
which property relies on the identity
$
\f{2z}{(1-z)}= -1+\f{1+z}{2-(1+z)}\,.
$
This allows us to use  $\widetilde G_\nu = G_{u_{1,-1/2}(\nu)}$ as the generating function
for elements of $M^\mathbf{T}(A)$. Composition law (\ref{diamaf}) gives
\be
\label{GGG}
\widetilde G_\nu\diamond \widetilde G_\mu = \widetilde G_{\gs_{1,-\half}(\nu,\mu)}\,.
\ee
Remarkably, $\gs_{1,-\half}(\nu,\mu)$ obeys (\ref{rhod}) provided that $\nu$ and $\mu$
do as one can easily see expanding $\gs_{1,-\half}(\nu,\mu)$ (\ref{fus}) in power series.
Hence, the composition law (\ref{diamaf}), (\ref{fus}) gives directly the composition law
in $M^\mathbf{T}(A)$. (A priory, it could happen that the composition of two generating
functions gives a generating function that does not respect  (\ref{rhod}),
hence requiring factorization of elements of the ideal that might complicate the problem
enormously.) Note that, in particular, these formulae provide a simple realization
of the universal enveloping algebras of orthogonal and symplectic Lie algebras since
the latter are subalgebras of $gl_n$ extracted by the antiautomorphisms $\rho$
generated by symmetric and antisymmetric bilinear forms, respectively.

In fact, the current operator algebra of \cite{Gelfond:2013xt} is associated with
 $M^\mathbf{T}(A)$ where the antiautomorphism $\rho$ of $A$ is defined as
$\rho f(Y^A) = i^{\pi(f)} f(iY^A)$. Indeed, it is well-known that
nontrivial conserved currents $J_s^{ij}$ of (odd)even  spins are (anti)symmetric
in their color indices $i,j$. As explained in \cite{Gelfond:2013xt}, this happens just
because they obey the condition
$
\rho (J^{ij}_s) = - J^{ij}_s.
$

\subsubsection{Ideals induced by central elements}
\label{cel}

Let $\cc_a$ be a basis of the centrum $C(A)$ of $A$, that forms  a subset of $t_i$.
In terms of  structure coefficients (\ref{struc}), this implies
\be
f^i_{aj} = f^i_{ja}\,.
\ee
By virtue of (\ref{circ}), elements $h(\cc,Id)\in M(A)$ are central in $M(A)$. Any
$h(\cc,Id)$ generates a two-sided ideal of $M(A)$.
In particular,  ideals $I_{c_a}$
\be
F(t)\in I_{c_a}:\quad F(t)=\prod_a ({\cal C}_a -c_a Id) \circ G(t)\q G(t)\subset M(A)\,,
\ee
as well as the quotient algebras
$
M_{c_a}(A)=M(A)/I_{c_a},
$
are parametrized by $c_a\in \mathbb{K}$.

A particularly important case is where $A$ is a unital algebra and $\cc=e_\star\in A$.
Then  $M_c(A)$ is parametrized by a single parameter $c$ resulting from factorization
of elements proportional to $e_\star-cId$
\be
\label{idId}
M_c:\quad e_\star-cId\sim 0\,.
\ee
In the case where trace is supported by $e_\star$,
the shift $e_\star\to e_\star +c Id$ gives the composition law
\be
\label{circtr}
F(\ga)\circ G(\ga) = F(\ga) \exp \Big (
\f{\overleftarrow{\p}}{\p \ga_i} (f^n_{ij} \ga_n +c g_{ij})
\f{\overrightarrow{\p}}{\p \ga_j} \Big ) G(\ga)\,.
\ee
 In the shifted variables,  $M_c(A)$ results from
dropping  $e_\star$ in all formulae.
As a result, Eq.~(\ref{circtr}) where
$F$ and $G$ depend only on traceless $\ga$,  describes the composition law in
$M_c(A)$.

In the context of
HS theories,  relevance of algebras $M_c(A)$ is not clear, however,
 since, as explained  in Section \ref{modul},
the factorization (\ref{idId})  identifies the physical vacuum
(no particles) with the lowest energy state of the space of single-particle
states.

\subsubsection{Finite-order quotients}
Within infinite zoo of ideals of $M(A)$ we will be particularly interested in
those that lead to quotient algebras realized by a finite number of tensor products
of $A$. Given function $\Phi(f_n, tr(f_m))$, the span of elements of the form
\be
\label{ideal}
\sum_\ga \Phi(f^\ga_{n}, tr(f_m^\ga)) \circ G^\ga\q \forall f^\ga\in A\,, G^\ga\in M(A)
\ee
forms a two-sided ideal $\II_\Phi$ of $M(A)$. Indeed,
since various  $h\in A$ generate $M(A)$ it suffices to show that
\be
\label{hff}
h\circ \II_\Phi \in \II_\Phi\q h\in A\,.
\ee
By virtue of (\ref{circ}) we observe
\be
h\circ \Phi(f_{n}, tr(f_m)) - (-1)^{\pi(h)\pi(\Phi)}\Phi(f_{n}, tr(f_m)) \circ h =\sum_{k} [h\,,f_{k}]_\star
\f{\p}{\p f_{k}} \Phi(f_{n}, tr(f_m))\,.
\ee
Using that $[h\,,\ldots]_\star$ is a derivation and that $tr([h\,,\ldots ]_\star )=0$,
we obtain
\be
\label{id}
h\circ \Phi(f_{n}, tr(f_m)) = (-1)^{\pi(h)\pi(\Phi)}\Phi(f_{n}, tr(f_m)) \circ h+
\f{\p}{\p \lambda} \Phi(f_{n}(\lambda), tr(f_m(\lambda)))\Big |_{\lambda=0}\,,
\ee
where
\be
f(\lambda) = f+\lambda [h\,,f]_\star\q
f_{n}(\lambda) = \underbrace{f(\lambda)\star\ldots \star f(\lambda)}_n \,.
\ee
 This proves (\ref{hff})
since the \rhs of Eq.~({\ref{id}) can be represented as a linear combination
of  polynomials $\Phi(f_{n}, tr(f_m))$ with different $f$.

Naively, factorization over order $n+1$ polynomials
$\Phi(f)=(f)^{n+1}+\ldots $ should give an algebra spanned by order $n$
polynomials. However, in most cases, this is not true because
the ideal $\II_\Phi $ turns out to be much larger, coinciding with $M^\prime (A)$.
Indeed, consider for example  bilinear $\Phi_\gamma(f_{n})$
\be
\label{fga}
\Phi_\gamma(f_{n}) = f^2 +2\gamma f\star f\,
\ee
with an arbitrary parameter $\gamma$. In this case, elements
\be
(fg +\gamma \{f\,,g\}_\star )\circ h = fgh +f(g\star h) + g(f\star h) +\gamma \{f\,,g\}_\star{}  h +
\gamma \{f\,,g\}_\star{}  \star{} h
\ee
belong to $\II_{\Phi_\gamma} $. Obviously,
\be
f(g\star{}h) + g(f\star{}h) \sim -\gamma (f\star{}g\star{}h + g\star{}h\star{}f +g\star{}f\star{}h +f\star{}h\star{}g)\,,
\ee
\be
\{f\,,g\}_\star{}  h \sim -\gamma (\{f\,,g\}_\star{} \star{} h + h\star{}\{f\,,g\}_\star{} )\,,
\ee
were equivalence is up to terms that belong to
$\II_{\Phi_\gamma} $.
This gives
\bee
(fg +\gamma \{f\,,g\}_\star{} )\circ h \sim&&  \ls fgh -\gamma (f\star{}g\star{}h + g\star{}f\star{}h +g\star{}h\star{}f +f\star{}h\star{}g)
 \\&&\ls-\gamma^2 (\{f\,,g\}_\star{} \star{} h + h\star{}\{f\,,g\}_\star{} ) +
\gamma \{f\,,g\}_\star{}  \star{} h\nn\\
&&\ls\ls = fgh -\gamma (g\star{}h\star{}f +f\star{}h\star{}g) -\gamma^2 (\{f\,,g\}_\star{} \star{} h + h\star{}\{f\,,g\}_\star{} )
\nn\,.
\eee
Antisymmetrization of this expression with respect to $h$ and $g$ gives
\be
\gamma(1-\gamma) [f,[h,g]_\star{}]_\star{}\in \II_{\Phi_\gamma} .
\ee
Hence, except for $\gamma=0$ or $\gamma=1$, $\II_{\Phi_\gamma}$ contains
all elements of $A$ that can be represented as $[f,[h,g]_\star{}]_\star{}$, \ie belong to
 the ideal $l_{(2)}(A)$ of $l(A)$. In the relevant cases where $l_{(1)}(A)$
($h\in l_{(1)}(A):$ $ h= [f,g]_\star{}$ for some $f,g\in l(A)$), and hence $l_{(2)}(A)$,
is simple, it coincides with almost all $l(A)$.
 Namely, in the cases of interest $l(A)=l_c(A)\oplus l_{(1)}(A)$ where $l_c(A)$ is the
Abelian algebra spanned by central elements of $A$. (This is fully analogous to
the relation $l(Mat_n(\mathbb{C})) = gl_n(\mathbb{C}) = sl_n(\mathbb{C})\oplus \mathbb{C}$,
where $l_{(1)}(Mat_n(\mathbb{C}))=sl_n(\mathbb{C})$.) Since $\II_{\Phi_\gamma}$ is too large
for generic $\gamma$,  we consider the special cases of $\gamma=0$ or $\gamma=1$.

Obviously, $M(A)/\II_{\Phi_0} = A\oplus \mathbb{K}$. On the other hand,
from (\ref{R2}) it follows that the case of $\gamma=1$ is related to $\gamma=0$ by
the principal antiautomorphism $\mathbf{R}$, \ie
 $M(A)/\II_{\Phi_{1}} = \tilde A\oplus \mathbb{K}$.
Hence the cases of $\gamma=0$ and $\gamma=1$ are exchanged via exchange of
$A$ with $\tilde A$
\be
M(A)/\II_{\Phi_\gamma}\sim M(\tilde A)/\II_{\Phi_{1-\gamma}}\q \gamma= 0,1\,.
\ee

As explained in  Section \ref{cel}, the unit elements  $e_\star\in A$
and $Id\in M(A)$ can be identified via factorization of the ideal $I_c$.
However, in the quotient algebras $M(A)/\II_{\Phi_\gamma}$ the parameter
 $c$ is no longer arbitrary. Indeed,
\be
e_\star\circ f= e_\star f + e_\star \star{}f \sim f-2 \gamma f =(1-2\gamma) f=(1-2\gamma)Id\circ f.
\ee
Hence, the factorization over both $I_c$ and $\II_{\Phi_\gamma}$ is possible at
$c(\gamma)= (1-2\gamma)$, \ie $c(0)=1$ and $c(1)=-1$.

The example of $\II_{\Phi_{0}}$ admits natural generalization to the ideals $\II^{N+1}$
generated by Eq.~(\ref{ideal}) with
\be
\label{fn1}
\Phi = f^{N+1}\,.
\ee
As a linear space, the quotient algebra $M_N(A):=M(A)/\II^{N+1}$ is
\be
M_N(A) = \sum_{n=0}^N \oplus Sym \,  \underbrace{A\otimes \ldots \otimes A}_n\,.
\ee
Here the only possible value of $c$ is
\be
c_N= N\,.
\ee
Indeed,
\be
e_\star \circ f^N = e_\star\, f^N + N (e_\star\star{}f) f^{N-1} \sim N f^N\,
\ee
since $e_\star\, f^N\in \II^{N+1}$.
Now one can consider quotient algebras
$
M^N(A)= M_N(A)/\II_{e_\star-N\,Id}.
$
Note that $M^1(A)=A$.
Similarly, the generalization of  $\II_{\Phi_1}$ to higher $N$
leads to ideals $ \II^{N+1}$ of $M(\tilde A)$ and quotients $M_N(\tilde A)$
and $M^N(\tilde A)$.

\subsection{Modules}
\label{modul} Since $M(A)\sim U(l(A))$, any $A$-module generates a
$M(A)$-module. Tensor product of any number of $A$-modules
forms a $l(A)$-module and, hence, $M(A)$-module. Beyond that, $M(A)$
admits less trivial modules which may be relevant in the
context of multiparticle HS theories.

Let $V$ be an $A$-module. Recall that in the HS context $A=H_{V_\Phi}$ and $V=V_\Phi$
is the space of single-particle states of some fields $\Phi$. The space of all
multiparticle states is
\be
\label{tensv}
\cV = \sum_{n=0}^\infty \oplus V^n\q V^n=
Sym \,  \underbrace{V\otimes \ldots \otimes V}_n\,.
\ee
When referring to a particular field theory  associated with $V_\Phi$ we
will use notation $\V_\Phi$.

Let $\tau_\ga$ be a basis of  $V$
\be
v\in V:\quad v=\sum_\ga v^\ga \tau_\ga
\ee
and
\be
\label{ttt}
t_i(\tau_\ga )= T_{i\ga}{}^\gb \tau_\gb\q (t_i \star{}t_j)(\tau_\ga)= T_{j\ga}
{}^\gga T_{i\gga}{}^\gb \tau_\gb\, \longrightarrow\,
 T_{j\ga}{}^\gga T_{i\gga}{}^\gb = f_{ij}^k T_{k\ga}{}^\gb\,.
\ee

Let $V^\star{}$ be dual to $V$
\be
\lambda\in V^*:\quad \lambda = \sum_\gb \lambda_\gb \tau^{*\gb}\,.
\ee
Similarly to the realization of $M(A)$ in terms of functions $F(\ga)$, elements of
$\cV$ can be represented by functions $\phi(\lambda)$ on $V^*$
\be
\phi(\lambda) = \sum_{n=0}^\infty \phi^{\ga_1\ldots \ga_n}\lambda_{\ga_1}\ldots \lambda_{\ga_n}
\,.
\ee
 Let  $F(\ga)\in M(A)$, $\phi(\lambda)\in \cV$.
 $\V$ can be endowed with the structure of $M(A)$-module by setting
\be
\label{cv}
F(\ga) (\phi (\lambda))= F({\ga})
\exp \Big (  \f{\overleftarrow{\p}}{\p { \ga}_i} t_i(\lambda_\gb)
\f{\overrightarrow{\p}}{\p \lambda_\gb} \Big ) \phi(\lambda)\Big |_{ \ga_i = {\mathbf t}_i(\lambda)}\,,
\ee
where ${\mathbf t}_i(\lambda)$ is some linear function of $\lambda_\ga$, that obeys the
condition
\be
\label{ttau}
t_i(\lambda_\gb) \f{\p {\mathbf t}_k(\lambda)}{\p \lambda_\gb}=f_{ik}^j{\mathbf t}_j(\lambda)\,.
\ee
That Eq.~(\ref{cv}) defines a $M(A)$-module,
\ie
\be
F(t)(G(t)(\phi(\lambda))= (F\circ G)(\phi(\lambda))\,,
\ee
is easy to see using Eqs.~(\ref{ttt}), (\ref{ttau}).

Let $\cV(V,{\mathbf t})$ be
the $M(A)$-module $\cV$ (\ref{cv}) determined by an $A$-module $V$ and
${\mathbf t}_i(\lambda)$  solving (\ref{ttau}).
The $M(A)$-module $\cV(V,0)$ is associated with  ${\mathbf t}_i=0$.
A less trivial option of
\be
\label{mt}
{\mathbf t}_i (\lambda)= o(t_i(\lambda))=o^{\ga} T_{i\ga}{}^\gb \lambda_\gb
\ee
is parametrized by a vector $o^{\gb}\in V$.
Indeed, in this case (\ref{ttau}) holds by virtue of (\ref{ttt}).

The module  $\cV(V,0)$ is infinitely reducible. Indeed, from Eq.~(\ref{cv}) with
${\mathbf t}_i=0$
it follows that the subspace $V^p$ of homogeneous polynomials $\phi(\lambda)$ of degree $p$ remains
invariant under the action of $M(A)$.
(Note that existence of the ideals $\II^{N+1}$ (\ref{fn1}) is closely related to the fact
of reducibility of $\cV(V,0)$: $\II^{N+1}$ is the annihilator  of $V^p$ with $p\leq N$.)
Clearly, $V^p$ are canonical $U(l(A))$-modules
associated with symmetrized tensor products of the $l(A)$-module $V$, \ie
spaces of $p$-particle states in their multiparticle interpretation.

For ${\mathbf t}_i\neq 0$ (\ref{mt}),
the action of $M(A)$ (\ref{cv}) is not homogeneous, mixing $V^n$ with different
$n$. Suppose that $V$ is induced from the vacuum vector $o$,
\ie $A(o)=V$. Let $V^*$ be  the right $A$-module and
$o^*\in V^*$ be normalized so that $o^*(o)=1$.
 From Eqs.(\ref{cv}), (\ref{mt}) it follows
that, in this case, the $M(A)$-module $\V(V,{\mathbf t})$ is induced from
the vacuum element $O\in \V$ identified with $\phi(\lambda)= 1$.
Modules of this type are somewhat analogous to Fock modules of oscillator
algebra (as is illustrated in the next section by the example of
Weyl algebra) and are expected to play an important role in
multiparticle theories.

$M_c(A)$ (\ref{idId}) results from $M(A)$ via factorization of
elements proportional to $e_\star-cId$. Similarly, a $M_c(A)$-module $\V_c(V,{\mathbf t})$
results from $\V(V,{\mathbf t})$ via factorization of elements induced from
\be
\label{oO}
o-c \,O\sim 0\,.
\ee
Indeed, according to (\ref{cv}), the action of of $e_\star-cId$ on $O$  gives
\be
(e_\star-cId )(O) = o(\lambda) -c \,O\,.
\ee
To reduce the $M(A)$-module $\V(V,{\mathbf t})$ to the
$M_c(A)$-module $\V_c(V,{\mathbf t})$ using relation (\ref{oO}),
 one has to remove the dependence on
$\lambda_o$ from $\lambda= \lambda_o o^* + \sum_\gb \tilde \lambda_\gb \tilde \tau^\gb $.
 In other words, $\V_c(V,{\mathbf t})$
consists of functions of all elements of $\V^*$ except for the vacuum $o^*$.
Since such a factorization  identifies the vacuum $o$ of the space of single-particle
states,  usually describing the lowest energy state of one or another
particle, with the physical vacuum $O$ with no particles, its physical meaning
is however obscure. In fact, the
difficulties of the naive extension of HS algebras discussed in Introduction
resulted just from consideration of $M_c(A)$-modules instead of $M(A)$-modules.

Another construction applicable to a Lie algebra $l(A)$ with any $A$,
which is particularly useful in the context of HS theory, is
that of twisted adjoint modules. Let $\tau$ be some automorphism of $A$.
The $\tau$-twisted adjoint $l_A$-module $A_\tau$ has $A$ as a linear space
where $l_A$ acts as follows
\be
a(b)=a \star b -b \star \tau (a)\q a\in l_A \q b\in A_\tau\,.
\ee
Any $\tau$-twisted adjoint module $A_\tau$ of $l(A)$ admits  straightforward extension
 to ${\mathcal T}$-twisted  adjoint module of $l(M(A))$. This simple observation is expected
 to play a key r\'{o}le for the formulation of a multiparticle generalization of HS gauge theory.

\subsection{Weyl algebra and Fock module}
\label{weyl}

 Weyl algebra $A_M$, which  underlies the construction of most of HS algebras,
 is the unital  algebra generated by
$2M$  elements  $Y_A$ satisfying (\ref{weyl1}). Remarkably, it
can itself be interpreted as the quotient of a multiparticle algebra $M(a_M)$. Here $a_M$ is
the algebra  with the generating elements $Y_\Omega$
and $h$ obeying  relations
\be
Y_A \star Y_B =K_{AB} h\q Y_A \star{} h=h\star{} Y_A =0\q
h\star{}h =0\,,
\ee
where $A,B=1,\ldots M$, and $K_{AB} $ is some matrix with the nondegenerate
antisymmetric part
\be
\label{CK}
C_{AB} = \big (K_{AB} -K_{BA}\big )\,.
\ee
Algebra $a_M$ is obviously associative since any triple product of
its  elements vanishes.

 $M(a_M)$ is spanned by functions $f(Y_A,h)$. This is not yet Weyl algebra, but
rather the algebra of quantum operators in
the deformation quantization framework with $h$ interpreted as a deformation
parameter. Weyl algebra $A_M =M_\hbar(a_M)$ results from $M(a_M)$ via
factorization of the ideal generated by $h-\hbar\, Id$ with parameter $\hbar$.
Note that  $M_\hbar(a_M)$ with various $\hbar\neq 0$
are pairwise isomorphic. The ``classical" case of $\hbar=0$ is
degenerate.

Different choices of the symmetric part of $K_{\Omega \Lambda}$
lead to different product laws (\ref{circ}) which correspond to different
star products for the same Weyl algebra.
Indeed, it is well known that different choices of
$K_{\Omega \Lambda}$ with the same $C_{\Omega \Lambda}$ (\ref{CK})
encode different ordering prescriptions.

Let us now explain how  Fock module of Weyl algebra
results from the construction of Section \ref{modul}. Consider for simplicity the case of
$a_1$ with the defining relations
\be
Y_- \star{}Y_+ = h\q Y_+ \star{}Y_-=0
\q Y_\pm \star{} Y_\pm=0\q Y_\pm \star{}h= h\star Y_\pm =0\q h\star{}h=0\,.
\ee
 A left $a_1$-module in a two-dimensional vector space $V$ with basis elements
$v$ and $v_+$, which can be realized as a quotient of the left adjoint
$a_1$-module, is
\be
Y_- v =0\q  Y_- v_+ = v\q
Y_+ v =0 \q  Y_+ v_+ = 0\q
h v=0\q h v_+ =0\,.
\ee
It is easy to see that
\be
\mathbf{t}_-(v) =0\q \mathbf{t}_+(v) =v_+\q \mathbf{t}_h(v) =v
\ee
solves Eq.~(\ref{ttau}). With this substitution, Eq.~(\ref{cv}) gives
the  $M(a_1)$-module realized by functions $\Phi(v_+, v)$ with
\be
\ga_- (\Phi) = v\f{\p}{\p v_+} \Phi\q \ga_+ (\Phi) = v_+ \Phi
\q \ga_h (\Phi) = v \Phi\,.
\ee
Factorization of the ideal of $M(a_1)$ generated by $h-\hbar\, Id$
along with its image in the constructed $M(a_1)$-module implies the
substitution $v\to \hbar$, giving the Fock module of the Weyl algebra $A_1$.

\section{Current operator algebra}
\label{Current operator algebra}

In this section we show how the current operator algebra in the twistor space
results from our construction.
 The space-time current operator algebra, which results from the twistor one via
 unfolded formulation of the current conservation equations, is
 not considered in this paper. We refer the reader to
 \cite{Gelfond:2013xt} on details of this relation.

The dictionary between notations of this paper and \cite{Gelfond:2013xt}
is as follows.
Free currents $\JJJ^2(U,V)$, where $U$ and $V$ denote twistor
variables used in \cite{Gelfond:2013xt}, identify with the generators
$t_i$ or, equivalently, with the basis $\ga_i$ of $A^*$.
The normal ordered product
$:\JJJ^2(U_1,V_1)\ldots\JJJ^2(U_n,V_n):$, which is by construction symmetric
with respect to the permutation of arguments $(U_b,V_b)$ with different $b$,
is represented by the  usual product $t_{i_1}\ldots t_{i_n}$.
Parameters $\nu^i$ have to be identified with
the parameters of currents called $g(W_1,W_2)$ in \cite{Gelfond:2013xt} and
usual powers $(\nu)^n$ represent $\JJJ^{2n}_g = :(\JJJ_g^2)^n:$.

The currents considered in \cite{Gelfond:2013xt} are invariant under
certain involutive operation $\mu$ as a consequence of
the construction of currents in terms of bilinears of free fields.
In the setup of this paper $\mu=-\rho$, where $\rho$ is the antiautomorphism
of Section \ref{genw}. The  current algebra
of \cite{Gelfond:2013xt} is nothing else but the quotient algebra
$M^{\mathbf T}(A)$ introduced in Section \ref{quan} where ${\mathbf T}$ is
the antiautomorphism of $M(A)$ generated by $\rho$.
As explained in Section \ref{quan}, the specific form
of the composition law associated with $\gs_{1,-\half}$ is compatible
with the (factorization) condition $\rho(\nu) = -\nu$ imposed in
\cite{Gelfond:2013xt}.

\subsection{$F$-current algebra}
\label{$F$-current algebra}

As mentioned in the end of Section \ref{quan},
$F$-current algebra at ${\mathcal N}=0$ is described by the composition law (\ref{GGG}).
To describe ${\mathcal N}$-dependent terms, one should generalize it using
(\ref{eta}) with appropriate function $\eta(\nu)$.
For
\be
\label{Trpp}
\eta(\nu)= Tr_{\Phi,\phi_\star}(G_{u_{1,\gb}(\nu)})\q \Phi(x) = \exp- x
\ee
with some $\phi_\star (\nu)$, the factor on the \rhs of (\ref{diamaftrc}) is
\be
\label{coc}
\exp tr \Big(\phi_\star \big (u_{1,\gb} (\nu) \star u_{1,\gb} (\mu) +u_{1,\gb} (\nu)
+ u_{1,\gb} (\mu) \big ) - \phi_\star \big ( u_{1,\gb} (\mu)\big )
- \phi_\star \big ( u_{1,\gb} (\nu)\big )\Big )\,.
\ee

Characteristic property of the $F$-current operator algebra
is that the trace-dependent part of the OPE for  both right and left
multiplication with the bilinear current $\JJJ_\mu^2$ only involves the trace
$tr(\nu\star \mu)$ between parameters of two elementary currents $\JJJ_\nu^2$
and $\JJJ_\mu^2$.  This imposes the condition  that
the part of (\ref{coc})  linear  either in $\mu$ or in $\nu$
should have the form $\f{1}{8}{\mathcal N} tr(\nu\star \mu)$. This
gives the differential equation
\be
\phi_\star'(u_{1,\gb}(\nu)) \star (u_{1,\gb}(\nu)+e_\star) = \f{1}{8}{\mathcal N} \nu
\ee
solved by
\be
\phi_\star (u_{1,\gb}(\nu))=  \f{1}{8}{\mathcal N}\left ( \gb^{-1} ln_\star (e_\star+\gb \nu)-
(1+\gb)^{-1} ln_\star (e_\star+ (1+\gb) \nu)\right )\,.
\ee
For $\gb=-\half$, this gives
\be
\phi_\star (u_{1,-\half}(\nu))= -\f{1}{4}{\mathcal N}\, ln_\star (e_\star-\f{1}{4}\nu\star \nu )\,.
\ee

Introducing
\be
\widetilde G_\nu = \eta (\nu) G_{{u_{1,-\half}(\nu)}},
\ee
the trace-dependent version of formula (\ref{diamaf}) at $b=1$,
$\gb=-1/2$ takes the form
\be
\label{F}
\widetilde G_\nu \diamond \widetilde G_\mu=
\left (\f{det_\star | e_\star-\f{1}{4}\nu\star \nu |\, det_\star | e_\star-\f{1}{4}\mu\star \mu |}
{det_\star | e_\star-\f{1}{4}\gs_{1,-\half}(\nu,\mu)\star \gs_{1,-\half}(\nu,\mu)|}
\right )^{\f{{\mathcal N}}{4}}
\widetilde G_{\gs_{1,-\half}(\nu,\mu)}\,,
\ee
where, as usual,
\be
\label{det}
det_\star |A|=\exp tr (ln_\star (A))\,.
\ee
(Of course, the $\star$-determinant possesses the
multiplication property $det_\star |A\star B| = det_\star |A|\,det_\star |B|$.)

Formula (\ref{F}) gives the generating function for the $F$-current operator
algebra of \cite{Gelfond:2013xt}. To see this it remains to check the parts
of $ \gs_{1,-\half}(\nu,\mu)$ linear either in $\nu$ or in $\mu$ which describe
left and right multiplication with $\JJJ^2_\nu$ and $\JJJ^2_\mu$, respectively.
Once, formula (\ref{F}) correctly reproduces this part of the
algebra, associativity implies that it describes the full
operator algebra.

For example, denoting by $\widetilde G^2_\mu$ the  part of $\widetilde G_\mu$ linear in $\mu$,
we obtain from (\ref{F})
\be
\widetilde G_{\nu} \diamond \widetilde G_{\mu}^2 =
\widetilde G_{\nu}\left (
\mu +\half (\nu\star \mu -\mu\star \nu ) -\f{1}{4}
\nu\star\mu\star\nu + \f{1}{8}{\mathcal N}\,tr(\nu\star \mu)Id \right ) \,.
\ee
These terms reproduce  OPE of $\JJJ_\nu^{2n}\,\JJJ_\mu^2$.
Indeed, the first term is the regular one. The second  results from
single contractions of the constituent fields. The third term results
from double contractions of the constituent fields of $\JJJ_\mu^2$
with two different  $\JJJ_\nu^2$ while the last one comes from
the double contraction of the constituent fields of $\JJJ_\mu^2$ with
those of some $\JJJ_\nu^2$.

Formula,
(\ref{F}) represents OPE of $\JJJ_\nu^{2n}\, \JJJ_\mu^{2m}$ by the
 $n$- and $m$-linear terms in $\nu$ and $\mu$. Note that
the ${\mathcal N}$-dependent central term does not contribute to
the commutator $\JJJ^2_\nu \JJJ^{2m}_\mu - \JJJ^{2m}_\mu \JJJ^2_\nu$.
This is because the dependence on ${\mathcal N}$ was introduced in
(\ref{Trpp}) in terms of a trace of $M(A)$.

\subsection{$A$-current algebra}

$A$-current algebra is the algebra of  currents with stripped
indices $a,b=+,-$ associated with the creation and annihilation parts of
the constituent fields. This algebra is anticipated to play a role in the
analysis of amplitudes (hence $A$-current algebra). Its structure differs
from that of $F$-current algebra in several respects. In particular,
the part of OPE associated with unity does contribute to the
operator commutator. Hence it cannot be derived via a field redefinition
(\ref{eta}) where $\eta$ is expressed in terms of some trace of $M(A)$ as
in (\ref{Trpp}), requiring $\eta$ of some other form.
Another novelty  is that  $A$-current algebra
 involves two associative products instead of one  in the
$F$-current case. This bi-associative structure underlies the construction
of butterfly product of \cite{Gelfond:2013xt}.
In this section, we first describe the relevant bi-associative structure
and then present explicit formulae for the  $A$-current operator algebra.

\subsubsection{Bi-associative algebra}
\label{biass}
By {\it bi-associative} algebra we
mean a linear space $A$ over a field ${\mathbb K}$ endowed with two mutually
associative products $\star$ and $\cdot$. This implies that $A$ is an associative
algebra with respect to the product
$
\ga \star +\gb \cdot
$
with any $\ga,\gb \in {\mathbb K}$. Equivalently,  in addition to
associativity of $\star$ and $\cdot$,
\be
(a\star b)\cdot c =a\star (b\cdot c)\q
(a\cdot b)\star c =a\cdot (b\star c)\,.
\ee

Biassociative algebras can be easily introduced as
follows.  Choose two elements $\tau_1$, $\tau_2$ of some associative algebra
$A$ with a product law $*$.  Then the two products
$a\star b := a*\tau_1*b$ and $a\cdot b :=a*\tau_2*b$ are associative
along with any their linear combination,  endowing $A$ with the
bi-associative structure.

$A$ is assumed to be unital  with respect to both $\star$ and $\cdot$.
However, the respective units $e_\star$ and $e_\cdot$ may be different.
They satisfy the following obvious relations
\be
e_\cdot \cdot e_\cdot = e_\cdot\q e_\star \star e_\star=e_\star\q
e_\cdot \cdot e_\star = e_\star \cdot e_\cdot =e_\star\q
e_\cdot \star e_\star = e_\star \star  e_\cdot  = e_\cdot\,.
\ee
One can see that, in addition, via relative rescaling of the two products
(and, hence, respective units) it is possible to achieve that
\be
\label{csc}
e_\cdot \star e_\cdot = e_\star\q e_\star \cdot e_\star = e_\cdot\,.
\ee
In the sequel, (\ref{csc}) is assumed to be true. These relations imply
 in particular
\be
\label{adb}
a\cdot b = a\star e_\cdot \star b\q
 a\star b = a\cdot e_\star\cdot  b
\ee
and that
\be
\label{pipm}
\Pi_{\pm}=\half (e_\star \pm e_\cdot )
\ee
are projectors with respect of each of the products
\be
\Pi_{\pm}\star \Pi_{\pm} = \Pi_{\pm}\cdot \Pi_{\pm}= \Pi_{\pm}\q
\Pi_{\pm}\star \Pi_{\mp} = \Pi_{\pm}\cdot \Pi_{\mp}= 0\,.
\ee

Inverse elements are defined  as usual
\be
a_\star^{-1} \star a =a \star a_\star^{-1} = e_\star\q
a_\cdot^{-1} \cdot a =a \cdot a_\cdot^{-1} = e_\cdot\,.
\ee
Property (\ref{csc}) has the following nice consequences
\be
a^{-1}_\star = e_\star \cdot a^{-1}_\cdot \cdot e_\star\q
a^{-1}_\cdot = e_\cdot \star a^{-1}_\star \star  e_\cdot\,,
\ee
\be
\label{inab}
(a\cdot b)^{-1}_\star = b^{-1}_\star \cdot a^{-1}_\star\q
(a\star b)^{-1}_\cdot = b^{-1}_\cdot\star a^{-1}_\cdot\,.
\ee
Note that from here it follows that $e_\star$ and $e_\cdot$ coincide with
their inverses with respect to each of the products
\be
(e_\star)^{-1}_\cdot = e_\star\q (e_\cdot)^{-1}_\star = e_\cdot\,.
\ee

In $A$-current operator algebra an important role is
played by an involutive  atiautomorphism $\rho$ obeying
\be
\label{rp}
\rho (a\star b) = \rho(b) \star \rho (a)\q \rho^2 = Id\q
\rho(e_\star)= e_\star \q \rho(e_\cdot)= -e_\cdot\q \rho (\Pi_\pm ) = \Pi_\mp \,.
\ee
By virtue of (\ref{adb}), this implies
\be
\label{ar}
\rho (a \cdot b) = -\rho (b)\cdot \rho(a)\,.
\ee

The products $\triangleright$ and $\triangleleft$ of
\cite{Gelfond:2013xt} are
\be
\label{><}
a \triangleright b = - a\star \Pi_+ \star b\q
a \triangleleft b = a\star \Pi_- \star b\,.
\ee

\subsubsection{Bi-maps and OPE}
Starting from the algebra $A$ endowed with the product law $\star$,
we can now apply the linear map (\ref{ubgb}) with $u(f)$ built with the
aid of $\cdot$. For maps analogous to (\ref{aff})
\be
\label{bimap}
 u_{b,\gb}(f) = b f \cdot (e_\cdot+\gb f)^{-1}_\cdot=\f{b}{\gb}
\left (e_\cdot-(e_\cdot+\gb f)^{-1}_\cdot\right )
\ee
an elementary computation gives again formula (\ref{diamaf}) where
 \be
\label{bisig}
\gs_{b,\gb}(\nu,\mu) = -\gb^{-1} \Big (e_\cdot-(e_\cdot+\gb \mu )\star
\big [e_\cdot-(\gb \nu- 2\Pi_-  )\star (e_\cdot +b\gb^{-1}e_\star )\star
(\beta \mu - 2\Pi_- )\big ]_\cdot^{-1}\star
(e_\cdot+\gb \nu )\Big )\,.
\ee
At $\cdot = \star$, $e_\cdot=e_\star$, this formula reproduces (\ref{sig}).
The map, that leads to the basis corresponding to the $A$-current OPE, is a
composition of some maps (\ref{munu}) and (\ref{bimap}).

Indeed, from general properties of the current operator
algebra it follows that it should be isomorphic to the algebra
 $M^{\mathbf T}(A)$
associated with an appropriate antiautomorphism $\rho$.
Hence, the product law of \cite{Gelfond:2013xt} should result from
(\ref{GGG}) via some field redefinition analogous to (\ref{ubgb}). However,
now we should confine ourselves to such field redefinitions that leave
$M^{\mathbf T}(A)$ invariant. This condition rules out
 field redefinitions (\ref{ubgb}), (\ref{u(f}) constructed in terms
of the  product $\star$
because even $\star$-degrees of elements $\nu$
obeying (\ref{rhod}) do not obey this condition, hence not belonging to
$M^{\mathbf T}(A)$. The trick is that it is
possible to use the product law $\cdot$, which  is $\rho$-odd itself
obeying (\ref{ar}), to perform a change of variables (\ref{bimap}) that maps
$M^{\mathbf T}(A)$ to itself.

Thus, the map $U^\star_{1,-\half}$ with respect
to the product $\star$ should be followed by some map  $U^\cdot_{b,\gb}$ with
respect to $\cdot$. It turns out that the appropriate form of the current operator
algebra results from the map $U^\cdot_{1,-\half}$. Practically, it is most
convenient to apply the map
\be
U^{\cdot\star}_{1,-\half}(G_\nu)= U^\cdot_{1,-\half}\big (U^\star_{1,-\half}(G_\nu)\big)
\ee
directly to (\ref{fnumu})
rather than to apply $U^\cdot_{1,-\half}$ to the product law (\ref{GGG}).

Using formulae of Section \ref{biass}, it is not difficult to obtain that
\be
U^{\cdot\star}_{1,-\half}(G_\nu) =
G_{\nu\star (e_\star -\Pi_+ \star \nu )^{-1}_\star}\,
\ee
with $\Pi_+$  (\ref{pipm}). The product law in the new basis
is
\be
\widetilde G_\nu * \widetilde G_\mu = \widetilde G_{\gs^{\cdot\star}_{1,-\half}(\nu,\mu)}\,,
\ee
where
\be
\gs^{\cdot\star}_{1,-\half} (\nu,\mu) = (e_\star +\nu\star \Pi_-)\star
(e_\star+\mu\star\Pi_+\star\nu\star\Pi_-)^{-1}_\star \star \mu +
(e_\star -\mu\star \Pi_+)\star(e_\star+\nu\star\Pi_-\star\mu\star\Pi_+)^{-1}_\star \star \nu\,.
\ee

To reproduce the trace-dependent terms we proceed as in the $F$-current
algebra case, requiring that if such terms are linear in $\mu$,  they
should  also be linear in $\nu$ having the form
\be
\label{atr}
-{\mathcal N}\,tr (\Pi_+\star \nu \star\Pi_-\star \mu)\,.
\ee
This is achieved via (\ref{diamaftrc}) with
\be
\tilde\eta ( x) = \exp [- {\mathcal N}\, tr  (ln_\star (e_\star-\Pi_+ x))]\,,
\ee
giving the composition law
\be
\label{A}
\widetilde G_{\nu} \diamond \widetilde G_{\mu}=
\left (\f{det_\star | e_\star-\Pi_+\star \gs^{\cdot\star}_{1,-\half}(\nu,\mu)|}
{det_\star | e_\star-\Pi_+ \star \nu |\, det_\star | e_\star-\Pi_+\star \mu |}
\right )^{{\mathcal N}}
\widetilde G_{\gs^{\cdot\star}_{1,-\half}(\nu,\mu)}\,.
\ee
This formula encodes in a concise form the full operator algebra of free
$3d$ conserved $A$-currents. The prefactor on its \rhs
provides the generating function for two-point functions
$\langle \JJJ_\nu^{2n} \JJJ^{2m}_\mu\rangle$.

To compare this product with the $A$-current operator algebra of \cite{Gelfond:2013xt},
it remains to consider the part linear in $\mu$
\be
\widetilde G_\nu \diamond \widetilde G_\mu^2 = \widetilde G_\nu\big (
\mu+\nu\star \Pi_-\star \mu - \mu\star \Pi_+\star \nu -\nu\star \Pi_-
\star \mu\star \Pi_+
\star \nu - {\mathcal N}\,tr (\Pi_+\star \nu\star \Pi_- \star\mu) \big )\,.
\ee
Using (\ref{><}), this formula just reproduces OPE for
$J_\nu^{2n} J^2_\mu$ of \cite{Gelfond:2013xt}.

\section{Multiparticle Lie (super)algebras and further extensions}
\label{hssa}

As explained in Section \ref{HSA}, HS algebras in their conformal interpretation describe
maximal symmetries of the space of single-particle states of  free conformal field theory
of some fields $\Phi$.
We propose  that a multiparticle extension of HS symmetry, that can serve as the
symmetry algebra of the space $\V_\Phi$ of all multiparticle states of  fields $\Phi$,
is provided by an appropriate real form of the Lie (super)algebra
 $l(M(H_{V_\Phi}))$. Specifically, we consider the Lie algebra
$\mmu(V_\Phi)$ which is the real form of $l(M(H_{V_\Phi}))$ singled out by the condition
\be
\label{S}
{\mathcal S} (f) = f\,,
\ee
where ${\mathcal S}$  is the conjugation of $l(M(H_{V_\Phi}))$ induced
via (\ref{sdag}) by the conjugation $\sigma$ in the reality
conditions (\ref{sa}) for $\hu(V_\Phi)$
(see \cite{Konstein:ij,Vasiliev:1999ba,Vasiliev:2003ev}).

Indeed, in accordance with Eq.~({\ref{cv}), $M(H_{V_\Phi})$ acts on
the space $\V_{\Phi}$ (\ref{tensv}) of all multiparticle states of
the field $\Phi$. Hence, the Lie algebra $l(M(H_{V_\Phi}))$ is the
complexified symmetry algebra of  $\V_\Phi$, while $\mmu(V_\Phi)$ represents
the appropriate real symmetry  of $\V_\Phi$.

Real algebras
$\mo(V_\Phi)$ and $\musp(V_\Phi)$ are  singled out by the
same condition (\ref{S}) from the complex Lie algebras
$M^\mathbf{T}(H_{V_\Phi})$ introduced in Section \ref{quan} by virtue of
the antiautomorphism $\rho$ used in (\ref{hou}) to single out subalgebras
$\ho (V_\Phi)$ or $\husp (V_\Phi)$ from $\hu(V_\Phi)$.

As a linear space,
\be
\label{V*V}
H_\V = \sum_{m,n=0}^\infty (V^m)^\star{} \otimes V^n\,,
\ee
while
\be
\label{MV}
\ls M(H_V) = \mmu(V) = \sum_{n=0}^\infty \oplus Sym \,  \underbrace{(V^\star\otimes V)
\otimes \ldots \otimes (V^\star\otimes V)}_n\,.
\ee
$\mo(V)$ and $\musp(V)$ are represented by
the further (anti)symmetrization $(V^\star\otimes V)_\rho$ of
$V^\star\otimes V$
\be
m_\rho(V)
= \sum_{n=0}^\infty \oplus Sym \,  \underbrace{(V^\star\otimes V)_\rho
\otimes \ldots \otimes (V^\star\otimes V)_\rho}_n
\ee
upon the identification
$
V^\star =\rho(V)
$
expressing the condition (\ref{rcol}).
Obviously,
 multiparticle extension of a HS algebra is not
itself a HS algebra.

It is useful to use the following realization of multiparticle algebras.
Consider $N$ copies of generators $t_i^\ga$ of $A$ ($\ga=1\ldots N$),
 that satisfy
\be
\label{ncop}
t^{\ga}_i \star{} t^\ga_j =f_{ij}^k t_k^\ga\q
t^\ga_i \star{}t^\gb_j = t^\gb_j \star{}t^\ga_i\quad \ga\neq \gb\,.
\ee
Consider the subalgebra of the enveloping algebra of these relations
spanned by  polynomials $P(t_i^\ga)$ that are symmetric under
 the group $S_N$ permuting different species  $t_i^\ga$, \ie
\be
P(t_i^\ga)\in M_N\,:\qquad T_{\ga\gb} P(t)= P(t) T_{\ga\gb}\,,
\ee
where $T_{\ga\gb}$ are generators of $S_N$, which exchange species $\ga$ and $\gb$
\be
T_{\ga\gb}= T_{\gb\ga} \q T_{\ga\gb} t^\gb_i= t_i^\ga T_{\ga\gb}
\ee
(no summation over repeated indices). This algebra is isomorphic to
$M_N(A)$ via the following identification
\bee
&&\ga_i\in M_N(A):\qquad\,\,\,\,\quad  \sum_{\delta=1}^N t_i^\delta \,,\nn\\&&
\ga_i \ga_j \in M_N(A):\qquad\quad  \half \sum_{\delta \neq \gb =1}^N t_i^\delta \star t_j^\gb  \,,\nn\\
&&\ga_i \ga_j \ga_k\in M_N(A):\qquad\, \f{1}{6}\sum_{\delta \neq \gb\neq \gga\neq
\delta =1}^N
t_i^\delta \star t_j^\gb \star t_k^\gga\,,
\eee
{\it etc}. Being isomorphic to $M_N(A)$ as a linear
space and having proper adjoint action of $l(A)$ generated by $t_i$,
the resulting algebra is isomorphic to $M_N(A)$. For $N=\infty$, this
construction gives  $M(A)$. Obviously, $M_1(A)=A\oplus \mathbb{K}$.

Algebras $M^N(A)$ result from $M_N(A)$ via factorization of the ideal (\ref{idId}).
The unit element $e_\star$ of $A$ is realized as $e_\star=\sum_\ga e_\star^\ga $.
In accordance with  Section \ref{ideals}, for $M_N(A)$ with finite $N$, the
only possible value of $c$ is $N$
since, for any element $f_N$ of maximal degree $N$,  $e_\star \star F_N = N F_N$.
As a result, the naive $N\to \infty$ limit of $M^N(A)$, via
extension of the number of species of $t_i^\ga$ to infinity may be
problematic leading to divergent $c$. Again, this indicates
that algebras $M_c(A)$ may have no direct physical application.

In fact, the difficulties with the extension of HS algebras discussed
in Introduction resulted just from the condition that all species
of oscillators  in (\ref{N}) have common unit element. In other words,
the difficulties were due to an attempt to use algebras $M^N(A)$. In the framework
of algebras $M_N(A)$ and $M(A)$, oscillators of any sort (index $\ga$)
have their own unit elements $e_\star^\ga$. Single-particle states are
realized as
$
N^{-1/2}\sum_{\ga=1}^N |\psi_\ga\rangle
$.
In particular, the lowest energy single-particle state is
represented by
\be
\label{sing}
N^{-1/2} \sum_{\ga=1}^N |0_\ga\rangle\,,
\ee
where $|0_\ga\rangle$,
represents the lowest energy state in the respective sector.
It suffices  to require
\be
e_\star^\gb |0_\ga\rangle = \delta^\gb_\ga |0_\ga\rangle
\ee
to achieve that the action of generators (\ref{N}) on the state (\ref{sing})
remains the same as in the original $N=1$ module,
hence escaping the problem with lowest energies.

The linear space of $M(H_V)$ (\ref{MV}) represents the space
of multiparticle states of the bulk HS theory which is different from the
space of multiparticle states  $\sum_{n=0}^\infty\oplus V^n$ of the boundary theory.
In field-theoretical terms,  algebra $M(H_V)$ and its
associated Lie (super)algebra are well suited for the description of
the bulk multiparticle theory that does not include boundary fields.
It may, however, be useful to unify both types of
fields in the same framework. For example, such a generalization should
underly the extension of the analysis of the $AdS_4/CFT_3$ HS correspondence of
\cite{Vasiliev:2012vf} to the case where boundary currents are built from
the boundary conformal fields. This can be achieved via the following generalization of
the proposed construction.

Consider algebra $H_{V'}$ spanned by elements
\be
F\in H_{V'} \,:\qquad F=(f,|v\rangle,\langle v|,\phi)\quad f\in H_V\,,\quad |v\rangle
\in V\,,\quad \langle v|\in V^* \,,\quad \phi\in {\mathbb K}
\ee
with the product law
\be
F_1*F_2 = (f_1\star f_2 + |v_1\rangle \langle v_2 | , f_1|v_2\rangle +\phi_2 |v_1\rangle ,
\langle v_1 | f_2+ \phi_1 \langle v_2 |, \phi_1\phi_2 +  \langle v_1 |v_2\rangle )\,.
\ee
$H_{V'}$  can be interpreted as the algebras of endomorphisms of the space
$
V'=V\oplus {\mathbb K}\,.
$

Supertrace of $H_V$ generates supertrace of $H_{V'}$
\be
str_* F = str_\star f + \phi
\ee
provided that the inner product $\langle |\rangle$ is defined via
\be
 \langle v_1 |v_2\rangle = (-1)^{\pi_1\pi_2} str_\star ( |v_2\rangle \langle v_1|)\,.
\ee
Note that such HS algebras were used in \cite{Vasiliev:1995sv} for
the description of $2d$ HS  gauge theory.

Clearly, $M(H_{V'})$ contains all combinations of multiparticle and
multiantiparticle states of the original
boundary theory where $V$ was the space of single-particle states.
Since single-particle boundary states can be interpreted as singletons,
the resulting construction is analogous to that of singleton strings
discussed in \cite{Engquist:2005yt,Engquist:2007pr}. It seems to be
most appropriate for  the analysis of multiparticle amplitudes in
the boundary theory.

Finally, multiparticle algebras  admit further
extension to brane-like symmetries via algebras $M^p(A)$ defined inductively
\be
M^{n+1} (A) = M(M^n (A))\,
\ee
with
$
M^0 (A) = A$, $M^1(A) = M(A)\,.
$
For the oscillator realization of $A$ by functions of oscillators $Y^A$,
elements of $M^p(A)$ are represented by functions $f(Y)$ of
$Y^A_{i_1 i_2\ldots i_p}$ endowed with $p$ copies of indices $i_k$
running from $0$ to $ \infty$, such that $f(Y)$ is symmetric with respect
to permutation of $Y^A_{i_1 i_2\ldots i_k \ldots i_p}$ for any  $k$.
As the variables $Y^{A}_{ i_1 i_2\ldots i_p}$ are reminiscent of the modes on a
$p$-dimensional surface, the algebras $M^p(A)$ are expected to be
related to brane-like theories.  Note that, in the HS setup,  the continuous
spectrum  difficulty in brane theory discovered in \cite{de Wit:1988ct}
is likely to be resolved in units of the background curvature.
The  $p$-brane algebras acting in hypothetical generalized membrane theories,
introduced analogously to $\mmu (V)$, $\mo(V)$ and $\musp(V)$,
we call $m^p_u (V)$, $m^p_o(V)$ and $m^p_{ usp}(V)$, respectively.

\section{Conclusion}
\label{conc}
Extensions of HS algebras suggested in this paper are anticipated to
underly multiparticle extensions of HS gauge theories containing
mixed symmetry fields associated with higher Regge trajectories of String Theory.
An important feature of multiparticle  algebras, that differs them from known HS
algebras, is that they are not Lie (super)algebras of endomorphisms of unitary modules
associated with one or another set of relativistic particles. This modification
makes it possible to avoid difficulties of the naive extension of construction
of HS algebras to  mixed symmetry HS fields.

Known HS algebras $\hu(V)$ can be realized as (matrix valued) Weyl algebra for some set
of oscillators $Y_A$ and unit element $e_\star$ or a quotient of some its subalgebra.
Multiparticle algebra $\mmu(V)$ is
realized by symmetric functions $f_n(\tilde Y_1,\ldots ,\tilde Y_n)$ of  any number
$n=0,1,2\ldots $ of variables $\tilde Y_\ga= (Y_{\ga A},e_{\star\ga})$.
The spaces $F^n$ of  functions $f_n (\tilde Y_1,\ldots \tilde Y_n)$ with various $n$ are reminiscent
of $n^{th}$  Regge trajectories in String Theory. A space-time symmetry algebra
$s$ belongs to both  $ \hu(V)$ and  $\mmu(V)$.
Following \cite{more}, the idea is to try to
formulate a multiparticle HS theory in terms of differential forms
$\Omega_n(\tilde Y_1,\ldots,\tilde Y_n)$ of various
degrees $p\geq 0$.

Gravitational field  is associated with the 1-forms
valued in $s$.
The space-time symmetry algebra $s\subset \hu(V)$
 can be extended to a larger finite-dimensional subalgebra of $\mmu(V)$.
For example, in the case where $H_V$ is Weyl algebra,
one can consider the algebra spanned by two types of bilinears
\be
\label{f1}
f_1(Y) = f_{1\,AB} Y^A \star Y^B\q
f_2(Y) =f_{2\,AB} Y^A Y^B\,,
\ee
where symmetrized tensor product is replaced by usual product.
Various $f_1(Y)$ span  $sp(2M)$ while $f_2(Y)$ extends
it to $sp(2M)\oplus sp(2M)$ where the  $sp(2M)$ spanned by $f_1(Y)$
is embedded diagonally. Hence there is more room for the choice of
background fields in $\mmu(V)$ than in~$\hu(V)$.

The problem of increase of vacuum energies mentioned in Introduction
does not occur because the $\hu(V)$-module
$V\otimes V$, which describes symmetric massless fields, belongs to the
$\hu(V)$-module $\V$ which is also a $\mmu(V)$-module.

In HS gauge theory, free massless fields are formulated in terms of
gauge 1-forms valued in the adjoint representation of the HS algebra
and 0-forms $C$ valued in the module often called
Weyl module since it contains Weyl tensor in the spin two sector along with
its HS generalizations. In the unfolded formulation,
all degrees of freedom in the system are represented by 0-forms. Hence,
Weyl module treated as a module of the space-time symmetry algebra $s$
is complex equivalent to a
unitary module of single-particle states in the system.

 Weyl module is realized as the twisted adjoint module
with respect to automorphism $\tau$ that changes a sign of translations
in the $AdS$ algebra, \ie
\be
\tau (L^{ab}) =L^{ab}\q \tau (P^{a}) =-P^{a}\,,
\ee
where $L^{ab}$ and $P^{a}$ are generators of Lorentz transformations and
$AdS$ translations, respectively. As explained in
Section \ref{auto}, $\tau$ induces ${\mathcal T}$-twisted adjoint $\mmu(V)$-module.
Let us call it $\C$. As a linear space it is isomorphic to the sum of all
symmetrized tensor products of $C$
\be
\label{tensp}
\C = \sum_{n=0}^\infty \oplus Sym \,  \underbrace{C\otimes \ldots \otimes C}_n\,.
\ee
As such, it should be complex equivalent to the space of all multiparticle
states of the $AdS$ HS theory (not to be confused with the
space of multiparticle states of the boundary conformal theory). This
implies that the  extension of the HS theory based on $\mmu(V)$
should describe all multiparticle
states of the original HS theory while $\mmu(V)$ is a symmetry
that acts on these states. Since $C$ is complex equivalent
to a unitary module of the $AdS_d$ algebra $s$, its symmetrized
tensor products and hence $\C$ also do. This suggests that the algebra $\mmu(V)$
respects the admissibility condition of \cite{KV0}.

To extend construction to the full nonlinear multiparticle HS theory it
is necessary to extend the construction of \cite{more} to various algebras
$\mmu(V)$ associated with HS algebras $\hu(V)$.
 Hopefully,  solution to this problem can drive us to new understanding
of a fundamental theory underlying both String Theory and HS theory.

Another application of the multiparticle algebras is that they are expected
to fix unambiguously the form of all correlators of conserved conformal currents
of all spins. Indeed, as shown in \cite{Maldacena:2011jn,Maldacena:2012sf}
the form of current operator algebra for conserved conformal HS currents, and
hence correlators, is unique for $d>2$. This fact has been used
 in \cite{Didenko:2012tv}, where connected parts of $n$-particle correlators
where found with the essential use of their covariance under  HS symmetry which
determines  each of them up to a factor. The multiparticle algebra
proposed in this paper relates $n$-particle correlators with different $n$.
 Hence, it should determine all $n$-particle correlators up to an overall
 coefficient.

 One consequence of the analysis of this paper is that the number of
  constituent conformal fields ${\mathcal N}$ is not an essential parameter of the
  multiparticle algebra. In other words, current operator algebras with
  different ${\mathcal N}$ are all equivalent, being associated with different
  basis choices in the same multiparticle algebra. This phenomenon is
  closely related to the enormous ambiguity in the choice of trace operation in
  the multiparticle algebra:  the same multiparticle algebra can
  give rise to inequivalent  ${\mathcal N}$-dependent $n$-point functions once
  the latter are expressed in terms of different ${\mathcal N}$-dependent
   trace operations. Surprisingly, conclusions of the analysis of multiparticle
   algebras in this paper differs essentially from what we used to in $2d$ conformal theory
   where ${\mathcal N}$ does contribute to the central charge and
   cannot be removed by a basis change since
   central extension of Virasoro algebra is nontrivial. The related point is that multiparticle algebras considered
   in this paper possess no nontrivial central extension.
   In fact, this raises an interesting problem of reformulation of
$2d$ conformal field theory within the scheme proposed in this paper.
At the present stage, this
problem is not quite straightforward since the unfolded machinery, which maps
the space-time description to the twistor one used in this paper, has not been
yet developed far enough for $2d$ conformal models (see, however,
\cite{Vasiliev:1995sv}).

Another interesting point is that, beyond a few distinguished bases
leading to operator algebra of
free currents with different ${\mathcal N}$, there exist infinitely many
bases where the form of current operator algebra and $n$-point functions
do not respect the Wick theorem. This raises a question whether
or not this opens a way towards construction of non-free theories. Indeed, once
the basis changes within $M(A)$ relate free theories with different ${\mathcal N}$,
which are not equivalent as field theories, more general basis changes,
most of which do not respect the Wick theorem, may generate non-free models.
(Recall that unfolded equations map operators in the twistor space to conserved
space-time currents independently of their construction in terms of free
fields; see \cite{Gelfond:2013xt}.)
The nontrivial part of the story is to check which of the resulting non-linear
theories are standard conformal theories in the sense that
stress tensor (\ie spin two current) has standard OPE with other primary
currents. We hope to consider this interesting issue elsewhere.

\section*{Acknowledgments}
I am grateful to O.Gelfond  for stimulating discussions and useful comments and to
P.Sundell for the correspondence. This research
was supported in part by RFBR Grant No 11-02-00814-a.

\end{document}